\documentclass[acmtog,nonacm,screen]{acmart}

\usepackage{amsmath}
\usepackage{booktabs}
\usepackage{tabularx}
\usepackage{colortbl}
\usepackage{placeins}
\usepackage{algorithm}
\usepackage{algpseudocode}
\usepackage{CJKutf8} % Chinese author comments under PDFLaTeX.

\floatname{algorithm}{Alg.}

\setcopyright{none}
\authorsaddresses{}
\renewcommand{\keywordsname}{Keywords}

\makeatletter
\let\@ACM@checkaffil\@empty
\let\hyxmp@parse@acmart\relax
\makeatother

\AtBeginDocument{%
  \fancypagestyle{standardpagestyle}{%
    \fancyhf{}
    \fancyfoot[C]{\footnotesize\thepage}

  }
  \fancypagestyle{firstpagestyle}{%
    \fancyhf{}
    \fancyfoot[C]{\footnotesize\thepage}

  }
  \pagestyle{standardpagestyle}
}
\hypersetup{hidelinks}
\AddToHook{cmd/maketitle/after}{\hypersetup{pdfcreator={LaTeX}}}

\definecolor{todocolor}{RGB}{150,45,20}

\definecolor{xiaoyucolor}{RGB}{30,90,180}

\newcommand{\OursPhysical}{95.32}
\newcommand{\OursVisual}{70.23}
\newcommand{\OursHumanPreference}{61.41}

\newcommand{\PhysicalLead}{6.72}
\newcommand{\VisualLead}{10.34}
\newcommand{\MeanPhysicalLead}{13.52}
\newcommand{\MeanVisualLead}{11.50}
\newcommand{\CardPhysicalGain}{3.98}
\newcommand{\CardVisualGain}{6.73}

\title[Text2Sim: Agentic Physics-Based Simulation Generation with Distilled Expertise]{Text2Sim: Agentic Physics-Based Simulation Generation with Distilled Expertise}

\author{Xiaoyu Xiong}
\affiliation{\institution{Tsinghua University}}
\affiliation{\institution{Carnegie Mellon University}}

\author{Tsun-Hsuan Wang}
\affiliation{\institution{Genesis AI}}

\author{Yi-Ling Qiao}
\affiliation{\institution{Genesis AI}}

\author{Tao Du}
\affiliation{\institution{Tsinghua University}}
\affiliation{\institution{Shanghai Qi Zhi Institute}}

\author{Minchen Li}
\affiliation{\institution{Carnegie Mellon University}}
\affiliation{\institution{Genesis AI}}

\renewcommand{\shortauthors}{Xiong et al.}

\begin{document}

\begin{abstract}
Creating diverse physical simulations remains labor-intensive because assets, layout, physical parameters, motion, control, and rendering must be designed and debugged jointly. We present \textit{Text2Sim}, a simulation-specialized agentic pipeline that converts a text-only request into an executable, editable dynamic case. Built on Genesis, Text2Sim uses a hierarchical agentic structure that combines a Planner with specialized Writers, asset-generation tools, and an independent Critic. Compact skills (Debug Cards) distilled from graphics demonstrations provide role-specific physical guidance for execution-based repair. We evaluate physical quality, visual quality, and human preference on 42 held-out prompts spanning rigid, articulated, deformable, and cloth phenomena, with a paper-level split between experience construction and evaluation. We design automatic physical and visual scorers to evaluate the quality of the results, and Text2Sim achieves higher scores than all four state-of-the-art baselines on both metrics. In blinded user studies with these baselines, significantly more participants prefer Text2Sim than prefer the baselines, which is consistent with the results from our automatic scorers. The pipeline also supports a broad range of downstream applications; we select dataset construction and extension to multimodal input as two representative examples.
We will release the code, the Debug Card library, and a dataset of generated cases, each pairing the text prompt and rendered video with the executable program, assets, physical parameters, controls, and recorded states.
\end{abstract}


\begin{teaserfigure}
  \centering
  \includegraphics[width=\textwidth]{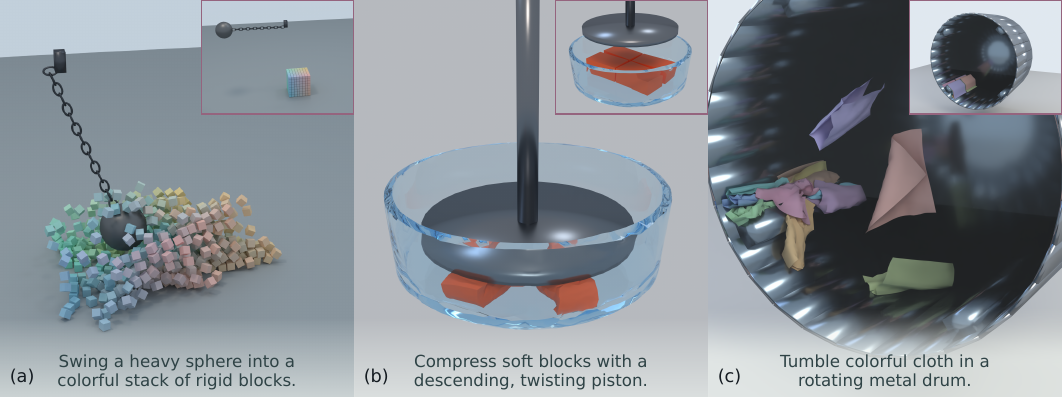}
  \caption{\textbf{Text2Sim generates physical simulations from text.} (a) A chain-suspended sphere scatters rigid blocks. (b) A piston compresses and twists soft blocks inside a cylindrical cup. (c) Colorful cloth pieces tumble inside a rotating drum. The embedded sentences summarize the input prompts. The large views show interaction snapshots from simulations lasting 10, 10, and 12\,s, respectively; upper-right insets show the initial frames.}
  \Description{Three edge-to-edge panels show rigid impact, soft-body compression, and cloth tumbling, with a short prompt centered near the bottom of each image and labels a, b, and c at the lower left. Panel a frames the complete chain-suspended sphere and scattered colored blocks at 1.44 seconds. Panel b shows orange blocks compressed and twisted beneath a piston in a transparent cup at 8.64 seconds. Panel c looks into a rotating dryer drum at 5.40 seconds, emphasizing the lifted, falling, and folded cloth. Equal-sized, softly bordered upper-right insets show the first recorded states without covering the main interaction: 0 seconds for panels a and b, and 2.04 seconds for panel c.}
  \label{fig:teaser}
\end{teaserfigure}

\maketitle

\section{Introduction}
\label{sec:introduction}

Authoring dynamic physical scenes for graphics and simulation is labor-intensive. Beyond geometry, this process involves defining physical behavior and configuring simulation and rendering. Experts traditionally assemble and repeatedly tune such scenes, making each new environment costly to develop; automation could lower the effort needed to explore a new physical event or produce diverse simulation data. Existing text-conditioned asset generators produce radiance fields, meshes, or point-based geometry~\citep{poole2022dreamfusion,chen2026sam3d,zhu2026relaxflow} for use as scene assets. Integrating these assets into a dynamic event requires additional scene-level configuration. Scene-generation systems such as Holodeck, SceneSmith, and SAGE~\citep{yang2024holodeck,pfaff2026scenesmith,xia2026sage} create plausible, simulation-ready layouts, with an emphasis on static composition. GenSim, RoboGen, and FactorSim~\citep{wang2024gensim,wang2024robogen,sun2024factorsim} automate environment generation for robotics and reinforcement learning, with representations and action spaces tailored to domain-specific objectives. General-purpose models such as GPT, Claude, DeepSeek, and Qwen~\citep{openai2023gpt4,hui2024qwen25coder} have strong agentic coding capabilities, although producing reliable simulation code can still be challenging. In particular, coordinating solver and control choices with rendering over long contexts can complicate generation and repair. A key challenge is therefore to coordinate these capabilities within a reliable dynamic simulation workflow.

We propose \textit{Text2Sim}, an agentic pipeline that transforms a text-only request into an executable, editable physical simulation (running on Genesis~\citep{Genesis}), acquiring the required assets without a user-provided asset library or database. Generating a dynamic event requires consistent choices about geometry, physical parameters, control, and rendering; reasoning about all of them within a single evolving context makes local failures difficult to isolate. We therefore design a multi-agent structure to resolve this, where a \emph{Planner} can maintain the global objective while specialized \emph{Writers} can work within explicit module interfaces, keeping each construction or repair task focused. Because a program can run successfully yet produce the wrong event, an independent \emph{Critic} examines both quantitative and qualitative execution evidence, and the \emph{Planner} routes its diagnoses to the responsible \emph{Writer} for targeted revision.

However, with this agentic structure, each agent is still built from general-purpose coding models and lacks simulation-specific skills to connect observed failures to appropriate physical checks and repairs. Hence, we obtain these skills by generating and debugging cases derived from 99 high-quality demonstrations from SIGGRAPH and SIGGRAPH Asia technical papers, distilling validated diagnoses and repairs into compact and general skills (\emph{Debug Cards}). The Planner retrieves cards according to the intended physical event, current execution evidence, and the receiving agent's role, so that guidance about contact, material choices, or actuation reaches the component responsible for the relevant decision. These cards provide in-context guidance without requiring each agent to process the full accumulated history. Together, coordinated authoring and retrieved experience support scenes spanning articulated rigid bodies, deformable solids, and cloth.

We evaluate Text2Sim on 42 held-out prompts: 10 rigid, 18 deformable, and 14 cloth cases, collected from published graphics demonstrations and cases designed by human simulation experts. To test generalization beyond the demonstrations used to construct Debug Cards, we enforce a paper-level split between card extraction and evaluation. The comparison includes GPT-5.6-Sol, Qwen-3.8-Max, Claude-Opus-5, and Code2Worlds~\citep{zhang2026code2worlds}. We use independent agent evaluators to assign two complementary, prompt-conditioned scores on a 0--100 scale (where higher scores are better). The physical score audits Scene, Body, Action, and Render requirements using code, runtime records, and rendered evidence; the visual score assesses content fulfillment, demonstration clarity, visual polish, and demo appeal from the prompt and video alone (Sec.~\ref{sec:evaluation_metrics}). Text2Sim achieves physical and visual scores of \OursPhysical{} and \OursVisual{}, exceeding the mean of the four baselines by \MeanPhysicalLead{} and \MeanVisualLead{} points, respectively. In a blinded user study, Text2Sim scores \OursHumanPreference{} in overall preference with GPT-5.6-Sol fixed at 50, and leads all four aggregate dimensions, which is consistent with the results from our automatic evaluators (Sec.~\ref{sec:user_study}). Ablation studies further show that the agentic structure and Debug Cards jointly contribute to the high-quality results of Text2Sim (Sec.~\ref{sec:ablation}).

Our pipeline also supports various downstream applications, including dataset construction and extension to multimodal input. The generated programs pair text and rendered video with explicit geometry, physical parameters, controls, and recorded states. These aligned data can be curated for vision--language model (VLM) training and evaluation or developed into simulator benchmark cases (Sec.~\ref{sec:dataset}). To support these uses, we will release the generated cases as a dataset, together with the code and the Debug Card library. For the extensions to multimodal input, we demonstrate two examples: a handwritten sketch specifies the shapes and order of colorful soft letters that fall, scatter, and settle into a pile; a supplied cloth mesh bearing an ACM SIGGRAPH texture is reused for two curtains that fold open and close (Sec.~\ref{sec:multimodal}). In both cases, the additional input specifies appearance or geometry, while text describes the desired dynamics.

In summary, this work makes the following contributions:
\begin{itemize}
  \item An authoring pipeline that maps open-ended text requests to executable and editable simulation code for rigid, articulated, deformable, and cloth cases, including asset preparation and rendering.
  \item A division of generation and repair among a Planner, specialized Writers, and an independent Critic, augmented with a retrievable Debug Card library of simulation-specific experience.
  \item An automatic evaluation protocol combining physical evidence and visual assessment, applied across three physical categories against four baselines and complemented by a blinded user study, ablations, and applications to dataset construction and multimodal input.
  \item A released dataset of generated cases spanning rigid, articulated, deformable, and cloth phenomena, each aligning text and video with executable code, assets, physical parameters, controls, and recorded states, together with the code and Debug Card library.
\end{itemize}

% Register the method overview early to balance the opening pages.
\begin{figure*}[!t]
    \centering
    \includegraphics[width=\textwidth]{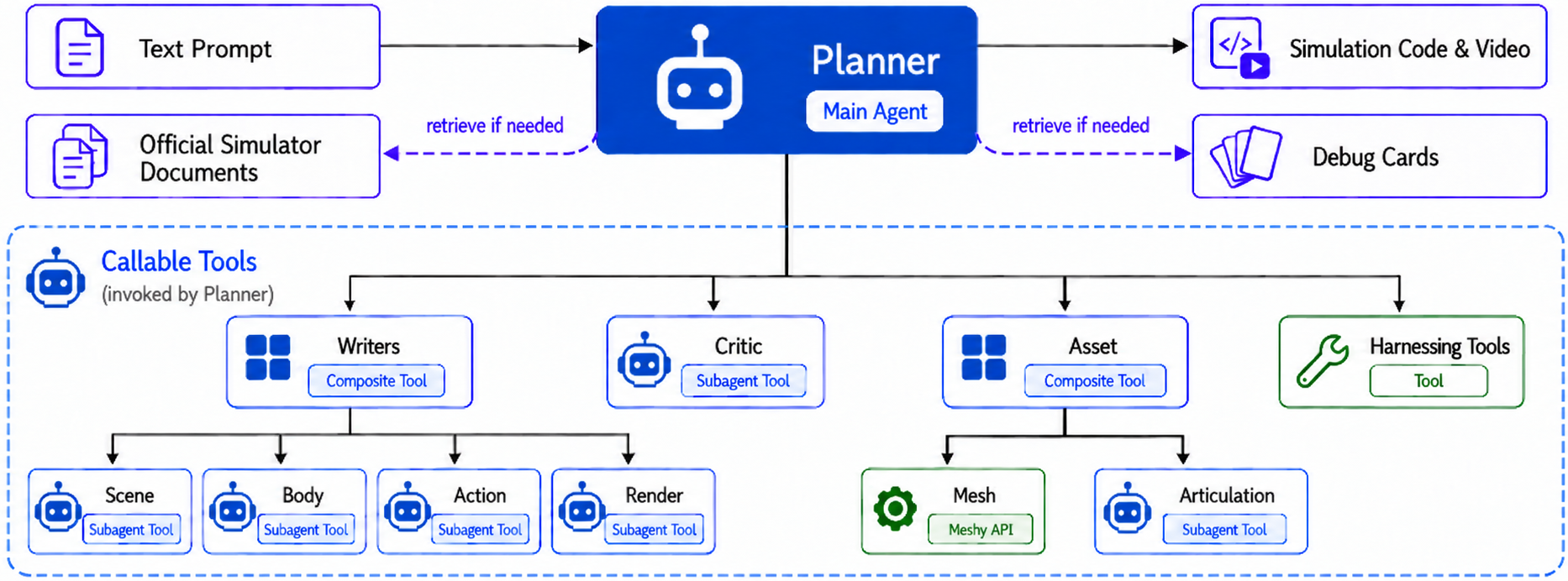}
    \caption{\textbf{Text2Sim pipeline.} The Planner receives a text prompt, accesses simulator documentation and Debug Cards (reusable simulation skills), and invokes the tools in the dashed box to produce simulation code and video. Four Writers author the Scene, Body, Action, and Render modules; the Critic evaluates executed cases; Mesh and Articulation agents provide assets; and harness tools execute the code. Solid arrows indicate control or artifact flow, and purple dashed arrows indicate on-demand retrieval. The antenna of the Planner indicates that it is responsible for sending commands, while the headphones of the subagents indicate that they are responsible for receiving commands. The two modules are colored green to indicate that, unlike the blue modules, they are not built on general-purpose LLMs.}
    \Description{The text prompt enters a Planner connected to simulator documentation and Debug Cards. The Planner invokes four Writers, a Critic, mesh and articulation asset tools, and an execution tool to produce code and video.}
    \label{fig:pipeline}
\end{figure*}

\section{Related Work}
\label{sec:related_work}

\paragraph{Physics-Based Simulation for Content Creation}
Physics-based simulation supports content creation by generating motion, deformation, and interactions. Traditional numerical approaches model continuum behavior using the material point method (MPM)~\citep{stomakhin2013snow,jiang2016material,liu2025ck}, smoothed particle hydrodynamics (SPH)~\citep{koschier2022survey}, and finite-element methods (FEM)~\citep{sifakis2012fem,li2026physics}. IPC and related contact formulations~\citep{li2020ipc,li2021codimensional,ferguson2021rigidipc,zheng2026barrierfree} handle contact and friction between objects. On the other hand, recent data-driven neural simulators offer another path by learning physical representations and motion models from simulated trajectories~\citep{chu2021controls,tao2024nirfs}. Across these approaches, constructing simulation scenes remains a substantial preparatory task. Benchmarking traditional solvers requires representative, reproducible test scenes; training neural simulators requires trajectories that cover suitable initial conditions and interactions. Both require specifying geometry, materials, initial and boundary conditions, and controls, then assembling and checking executable scenes. Text2Sim addresses this shared scene-authoring burden in solid mechanics by generating editable simulation programs from text, together with physical records that can support benchmark and training-data construction.

\paragraph{End-to-End Video Generation of Physical Processes}
End-to-end learned video generation offers a direct route for generating physical processes, including object motion, deformation, and contact. Models such as Video Diffusion Models~\citep{ho2022videodiffusion}, Lumiere~\citep{bar2024lumiere}, and Seedance 1.0~\citep{gao2025seedance} synthesize visual sequences conditioned on text or images. On the other hand, Text2Sim explores an alternative route to generating such physical processes by authoring and executing physical scenes. This route allows scenes to be inspected, modified, and rerun. As a consequence, our method is more compatible with applications that have strict physical constraints and require editable programs, physical parameters, controls, and recorded states alongside rendered video, such as robotics and engineering environments.

\paragraph{Domain-Specific Physical Scene Generation}
Scene creation involves building assets, arranging them in space, and specifying behavior when needed. WordsEye~\citep{coyne2001wordseye} maps text to static scenes, while relationship templates~\citep{zhao2016relationship} preserve spatial relations across scene variations. Image- or text-conditioned methods generate or reconstruct objects and rooms~\citep{poole2022dreamfusion,hollein2023text2room,chen2026sam3d,zhu2026relaxflow}; WorldExplorer~\citep{schneider2025worldexplorer} constructs navigable 3D Gaussian scenes. Holodeck, SceneSmith, SAGE, and PAT3D~\citep{yang2024holodeck,pfaff2026scenesmith,xia2026sage,lin2026pat3d} provide simulation-ready layouts. Specialized systems generate robotics or game environments~\citep{wang2024gensim,wang2024robogen,sun2024factorsim}, synthesize robot policies~\citep{liang2023codepolicies}, infer rewards from video~\citep{shi2025points2reward}, and use language and trained policies to control simulated humanoids~\citep{juravsky2022padl}. BrickGPT~\citep{pun2025brickgpt} and Learn2Fold~\citep{huang2026learn2fold} target stable brick assemblies and valid origami, respectively. These methods focus on geometry, initial layouts, or task-specific representations. Text2Sim, in contrast, aims to jointly author these components, including assets, physical parameters, and temporal behavior, to construct complete scenes across various physical environments.

\paragraph{Existing LLM-based Agentic Systems}
LLM-based agents can translate language requests into executable workflows through code generation and tool use. General-purpose LLMs such as GPT-4 and Qwen2.5-Coder~\citep{openai2023gpt4,hui2024qwen25coder}, together with Claude and DeepSeek, provide a foundation for code generation; ReAct~\citep{yao2022react} couples reasoning with actions, while SWE-agent~\citep{yang2024sweagent} uses execution feedback to repair programs. However, software execution alone cannot establish physical stability or confirm that the intended visual event has occurred. Graphics authoring introduces specialized representations: ParSEL~\citep{ganeshan2024parsel} combines LLM edits with analytical propagation to preserve part relations; \citet{goel2024motionediting} translate language requests into executable kinematic motion-editing programs; and \citet{gumin2025procedural} repair layouts through local program search, reducing spatial violations while preserving structure. CueTip~\citep{memery2025cuetip} grounds pool coaching in simulator-generated event traces and expert rules, illustrating how domain-specific feedback supports physical reasoning. Authoring complete scenes also requires coordinating geometry, materials, and temporal behavior. MCP-SIM~\citep{park2026mcpsim} coordinates multiple agents to generate scientific physics programs, ChronoLLM~\citep{wang2026chronollm} specializes in PyChrono, and Code2Worlds~\citep{zhang2026code2worlds} uses visual scene repair. SimuScene~\citep{wang2026simuscene} shows that reliably generating executable physical scenarios remains challenging even in 2D. The recent GS-Agent system~\citep{zhang2026gsagent} coordinates specialized agents and multimodal feedback within a physics-engine loop to generate dynamic worlds. Text2Sim further extends this direction with a purpose-designed agentic architecture that uses both quantitative and qualitative records, alongside reusable Debug Cards distilled from prior simulation examples, yielding high-quality results with greater robustness.

\section{Design Goals and Simulation Foundation}
\label{sec:design_goals}

Text2Sim targets authors who can describe a physical event but would otherwise need to assemble and debug its assets, simulation code, and visual presentation manually. Four goals organize the system described in Sec.~\ref{sec:method}.

\paragraph{Executable and editable output.}
The output should expose geometry, material parameters, motion and control, and rendering in a simulation program. This representation allows a user to inspect the construction of an event and revise its components. Preserving the program and its execution evidence also makes generated cases reusable for the dataset applications in Sec.~\ref{sec:dataset}.

\paragraph{Coverage across general solid scenes.}
One authoring interface should accommodate articulated rigid bodies, volumetric deformables, and cloth, including supported contacts between these families. Fig.~\ref{fig:teaser} illustrates this scope through rigid impact, compression and shear of soft blocks, and cloth--rigid contact.

\paragraph{Coordinated agent cooperation.}
Asset preparation, scene construction, physical parameters, control, and rendering require different expertise but must remain consistent. A Planner coordinates specialized Writers and an independent Critic, allowing each agent to focus on a bounded task while execution feedback guides targeted revisions. The main comparisons with the state-of-the-art baselines (Sec.~\ref{sec:main_results}) support the effectiveness of this coordinated design as a complete system.

\paragraph{Reusable physical experience.}
We extract reusable physical skills from prior SIGGRAPH demonstrations and encode them as skills (Debug Cards), which are retrieved to guide physical setup and repair. The ablation study (Sec.~\ref{sec:ablation}) supports the value of incorporating this experience into the pipeline.

% \section{Simulation Background}
\\
\label{sec:simulation_environment}

To support the generation of general solid-mechanics cases, we adopt Genesis~\citep{Genesis} as the underlying simulation environment and Meshy\footnote{\url{https://www.meshy.ai/}} as the mesh generator. After the agents construct a case and specify its motion or control commands, Genesis executes the resulting program and computes the physical evolution. The engine provides simulation models for articulated rigid bodies, volumetric deformable bodies, and cloth. In addition, its Incremental Potential Contact (IPC) coupler~\citep{li2020ipc} supports realistic two-way contact coupling between participants. Genesis also provides a path-tracing renderer based on LuisaRender~\citep{zheng2022luisarender} for producing high-quality visual observations of the simulated motion. These existing simulation and rendering capabilities support the object classes targeted by Text2Sim. Our method addresses the authoring decisions that configure them: geometry, physical parameters, controls, and presentation.

\section{Method}
\label{sec:method}

% Register the Writer illustration with the method overview.
\begin{figure}[t]
    \centering
    \includegraphics[width=\linewidth]{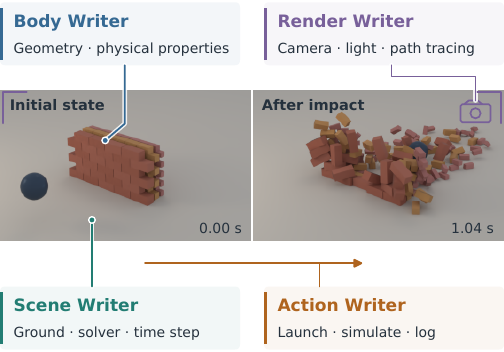}
    \caption{Writer responsibilities in a projectile--brick wall example, shown before and after impact. Green links the Scene Writer to the ground and simulation settings; blue links the Body Writer to object geometry and physical properties; orange denotes the Action Writer's launch, simulation, and logging; purple denotes the Render Writer's camera and lighting. The two path-traced frames share a crop, with simulation times labeled in seconds.}
    \Description{Two frames show an intact brick wall and a projectile at 0 seconds, followed by scattered bricks at 1.04 seconds. Blue connects the Body Writer to the wall, green connects the Scene Writer to the ground, orange links the Action Writer to the progression between states, and purple links the Render Writer to the image frame and a camera symbol.}
    \label{fig:writer_roles}
\end{figure}

\subsection{Problem and Overview}
\label{sec:problem}

The input is a text request describing objects, physical interactions, any actuation, and observable outcomes. The output is an executable simulation program with explicit assets, physical parameters, controls, and rendering settings, together with evidence of the simulated event. A useful output must both run and realize the requested behavior.

A Planner decomposes the request and coordinates generation and repair (Sec.~\ref{sec:planner}). Four Writers construct the static scene, movable bodies, actions, and rendering code (Sec.~\ref{sec:writers}); an independent Critic examines execution evidence and identifies needed revisions (Sec.~\ref{sec:critic}). Mesh and Articulation agents supply validated geometry and articulated mechanisms when required (Sec.~\ref{sec:asset_agents}). Debug Cards are compact records of simulation-specific diagnostic and repair guidance distilled from graphics demonstrations; the Planner retrieves relevant cards for each task and agent (Sec.~\ref{sec:debug_cards}). Fig.~\ref{fig:pipeline} shows these components, and Alg.~\ref{alg:procedure} summarizes the generation and repair loop.

\begin{algorithm}[!htb]
  \caption{Text2Sim generation and repair}
  \label{alg:procedure}
  \small
  \begin{algorithmic}[1]
    \Require Text request; attempt budget $B$
    \Ensure Simulation program, assets, and execution evidence
    \State Plan the case; initialize diagnosis $d \gets \varnothing$
    \For{$k = 1, \ldots, B$}
      \State Retrieve cards for the plan and $d$
      \State Dispatch agents to create or repair code and assets
      \State Integrate and check; execute and retain evidence
      \State Critic reviews the case and updates $d$
      \If{the Critic accepts the case}
        \State \Return program, assets, and evidence
      \EndIf
    \EndFor
    \State \Return best available attempt and unresolved failures
  \end{algorithmic}
\end{algorithm}

\subsection{Multi-Agent Structure}
\label{sec:multi_agent_structure}

In order to extend the capabilities of a single agent beyond its limited context, many prior works have designed agentic structures which combine reasoning, tool use, and execution feedback to support collaboration~\citep{yao2022react, wu2023autogen, yang2024sweagent}. Text2Sim specializes this pattern for simulation authoring, where geometry, materials, contact, control, numerical settings, and rendering must jointly realize a physical event. Our design separates source ownership while sharing explicit asset and control interfaces, so that a local revision can preserve the rest of the scene. Inspired by the separation of acting and evaluation in actor--critic methods~\citep{konda1999actorcritic}, the Writers construct the executable case, whereas an independent Critic evaluates its physical behavior using both numerical and visual evidence. A Planner maintains the global objective, coordinates code and asset revisions, and supplies relevant Debug Cards. This design links execution evidence to the relevant source owner and diagnostic guidance.

\subsubsection{Planner}
\label{sec:planner}

The Planner is the control-plane agent of Text2Sim. It does not directly author or review the generated source code. Instead, it decomposes the user request, dispatches well-scoped tasks to the downstream agents, receives their structured reports, and determines the next operation. It is also the only agent that communicates with the environment outside a case workspace. User instructions, simulator documentation, Debug Cards, execution permissions, and hardware resources are all mediated by the Planner; the other agents receive only the information and capabilities required by their assigned tasks. This boundary prevents a local code-generation task from silently expanding the scope of the overall pipeline.

For each case, the Planner first converts the text prompt into a global plan covering the intended physical event, simulation mode, timing, required assets, and module dependencies. It then consults the relevant Genesis documentation and local implementation when an API or solver choice must be resolved. Based on the resulting plan, it requests any necessary assets and invokes the appropriate Writers, which may run concurrently when their dependencies permit. Once a complete program is available, the Planner invokes the execution harness, for example to run Genesis on a local GPU, and forwards the resulting code, numerical records, and visualizations to the Critic. It uses the Critic's diagnosis to dispatch targeted revisions and repeats this process. The Planner owns the termination decision: it completes the case after the required execution and evaluation checks pass, or returns the best available result and failure evidence when the iteration budget is exhausted.

\subsubsection{Writers}
\label{sec:writers}

The Writers construct the executable simulation code. We separate the stage, actors, temporal behavior, and cinematography because each requires distinct information and exposes different repair controls. Their fixed interfaces and file ownership keep each context focused and make revisions local. Fig.~\ref{fig:writer_roles} illustrates this division using a projectile--brick wall example.

\paragraph{Scene Writer.}
The Scene Writer builds the \emph{stage} of the case. It initializes Genesis, selects global solver and contact options, sets time-stepping hyperparameters, and constructs static background geometry such as the ground, supports, or fixed boundaries. It does not create objects that move during the task or specify their controls.

\paragraph{Body Writer.}
The Body Writer creates the \emph{actors} of the case. It instantiates all movable rigid, articulated, deformable, and cloth objects and assigns their geometric properties, including scale and initial pose, as well as physical properties such as density, friction, and Young's modulus. It also exposes stable object identities and control handles so that the other modules do not need to reconstruct the semantics of an imported asset.

\paragraph{Action Writer.}
The Action Writer specifies the \emph{script} of the case. After initialization, it advances the simulation and applies the requested time-dependent operations, such as delivering an impulse to an object or sending a target signal to an actuator. Since many failures are only observable during execution, this module also inserts case-specific measurements and event logging. These records expose object trajectories, controller phases, contact events, and success-related quantities to the Planner and Critic without requiring them to infer all behavior from the rendered video.

\paragraph{Render Writer.}
The Render Writer controls the \emph{cinematography} of the case. It specifies lighting, camera extrinsics and intrinsics, frame-capture cadence, and video composition. Its interface is deliberately separated from physical state and control: visual adjustments can improve the legibility of an event, but cannot manufacture a successful physical outcome.

In the example, the Scene Writer supplies the fixed floor and global simulation settings, and the Body Writer constructs the separate rigid bricks and projectile. The Action Writer initializes the projectile velocity once, then advances the solver and logs the impact; the ensuing brick motion emerges from rigid-body contact. The Render Writer configures the camera and light placement and captures the cached states using path tracing.

After the four modules are produced, a deterministic integrator assembles them and checks their required interfaces. The ownership boundaries also define how repairs are applied. A camera occlusion is returned to the Render Writer, whereas an incorrect actuator schedule is returned to the Action Writer. Thus, an agent revises only the part of the case for which it has sufficient context and authority.

% Register the physical and visual tables together before the large comparison figures.
\begin{table*}[!t]
  \caption{Physical scores for our method and four baselines over all 42 cases (Total). Entries report the original means with their associated errors; bold marks the largest mean (best result) in each metric.}
  \label{tab:main_results}
  \label{tab:physical_results}
  \centering
  \fontsize{9}{10.5}\selectfont
  \setlength{\tabcolsep}{3pt}
  \begin{tabular*}{\linewidth}{@{\extracolsep{\fill}}lrrrrr@{}}
\toprule
Configuration & Scene & Body & Action & Render & Overall \\
\midrule
% Main-experiment Total results; 25 mean/error pairs per table.
Text2Sim & $\mathbf{99.10}\,{\pm}\,2.42$ & $\mathbf{99.15}\,{\pm}\,1.20$ & $\mathbf{96.10}\,{\pm}\,2.36$ & $\mathbf{87.22}\,{\pm}\,3.96$ & $\mathbf{95.32}\,{\pm}\,1.37$ \\
GPT-5.6-Sol & $99.09\,{\pm}\,3.08$ & $95.55\,{\pm}\,4.15$ & $82.69\,{\pm}\,4.91$ & $76.50\,{\pm}\,5.20$ & $86.59\,{\pm}\,4.44$ \\
Qwen-3.8-Max & $97.25\,{\pm}\,3.42$ & $93.38\,{\pm}\,4.22$ & $83.57\,{\pm}\,3.96$ & $70.89\,{\pm}\,4.40$ & $84.09\,{\pm}\,2.05$ \\
Claude-Opus-5 & $97.03\,{\pm}\,4.38$ & $96.84\,{\pm}\,2.69$ & $88.75\,{\pm}\,3.16$ & $77.63\,{\pm}\,4.40$ & $88.60\,{\pm}\,2.07$ \\
Code2Worlds & $67.20\,{\pm}\,7.04$ & $81.50\,{\pm}\,6.93$ & $71.68\,{\pm}\,6.09$ & $73.56\,{\pm}\,6.00$ & $67.91\,{\pm}\,4.27$ \\
\bottomrule
\end{tabular*}

\end{table*}

\begin{table*}[t]
  \caption{Visual scores for our method and four baselines over all 42 cases (Total). Entries report the original means with their associated errors; bold marks the largest mean (best result) in each metric. Content, Clarity, Polish, and Appeal denote content fulfillment, demonstration clarity, visual polish, and demo appeal, respectively.}
  \label{tab:visual_results}
  \centering
  \fontsize{9}{10.5}\selectfont
  \setlength{\tabcolsep}{3pt}
  \begin{tabular*}{\linewidth}{@{\extracolsep{\fill}}lrrrrr@{}}
\toprule
Configuration & Content & Clarity & Polish & Appeal & Overall \\
\midrule
% Main-experiment Total results; 25 mean/error pairs per table.
Text2Sim & $\mathbf{71.58}\,{\pm}\,4.82$ & $\mathbf{74.94}\,{\pm}\,3.95$ & $\mathbf{66.27}\,{\pm}\,2.73$ & $\mathbf{64.70}\,{\pm}\,4.75$ & $\mathbf{70.23}\,{\pm}\,2.87$ \\
GPT-5.6-Sol & $63.00\,{\pm}\,4.28$ & $64.98\,{\pm}\,3.23$ & $55.08\,{\pm}\,3.95$ & $51.48\,{\pm}\,4.30$ & $59.89\,{\pm}\,2.58$ \\
Qwen-3.8-Max & $60.10\,{\pm}\,3.86$ & $60.56\,{\pm}\,3.63$ & $52.86\,{\pm}\,4.13$ & $45.96\,{\pm}\,3.47$ & $56.31\,{\pm}\,2.51$ \\
Claude-Opus-5 & $62.44\,{\pm}\,4.11$ & $64.38\,{\pm}\,3.53$ & $54.89\,{\pm}\,4.02$ & $51.21\,{\pm}\,3.86$ & $59.45\,{\pm}\,2.61$ \\
Code2Worlds & $59.70\,{\pm}\,5.11$ & $60.52\,{\pm}\,3.63$ & $57.84\,{\pm}\,3.85$ & $58.27\,{\pm}\,3.63$ & $59.26\,{\pm}\,2.91$ \\
\bottomrule
\end{tabular*}

\end{table*}

\subsubsection{Critic}
\label{sec:critic}

The Critic evaluates physical behavior rather than execution success alone. It examines each complete intermediate version after execution, independently of Writer calls and with read-only access to the case. The Planner supplies the original text prompt together with the generated modules, asset manifests, execution logs, numerical metrics, event records, sampled frames, and video. Evaluating these sources jointly is important: a program may execute without error while producing the wrong event, and a visually plausible video may conceal non-causal state changes or invalid physical parameters.

The Critic returns both an overall prompt-alignment score and a discrete verdict indicating whether the case passes, requires repair, or cannot be judged from the available evidence. When repair is required, it identifies concrete failure modes, cites the supporting numerical or visual evidence, recommends the responsible module, and provides a specific revision summary. The Planner converts this report into a new, narrowly scoped request for the corresponding Writer or Asset agent. This evaluation loop separates generation from judgment and prevents a Writer from accepting its own output solely because the code runs.

\subsubsection{Assets}
\label{sec:asset_agents}

Many prompts require digital assets more complex than spheres, cubes, or other simulator primitives. Generating and validating such assets involves different tools and failure modes from writing the case program, so we implement an Asset branch independent of the Writers. When the Planner detects an asset requirement, it sends a concise textual description to the corresponding Asset agent. A validated result is registered in a shared asset manifest and is then exposed to the Writers through a stable path and interface.

We support two asset families. For a non-articulated object, the Mesh agent calls the Meshy\footnote{\url{https://www.meshy.ai/}} text-to-3D service to obtain geometry and appearance. The returned mesh is repaired and checked for topology, orientation, scale, and Genesis import compatibility before it can be used in a case. For an articulated object, a dedicated coding agent generates an MJCF/XML configuration that defines its body hierarchy, joints, collision geometry, and actuators. The mechanism is then checked structurally, imported into the simulator, previewed, and---when actuated---tested with a small control probe. In both families, the manifest records the resulting geometry and control contract, allowing the Body and Action Writers to consume an asset without repeating its generation logic.

% Separate the user-study table from the full-width comparison figures.
\begin{table*}[!t]
  \centering
  \caption{User-study scores for the five main methods over all 42 cases (Total). Entries report the original means with their associated errors; GPT-5.6-Sol is a fixed reference at 50. Bold marks the largest mean (best result) for each dimension. Cell backgrounds encode the mean relative to the reference: white at 50, increasingly green above 50, and increasingly red below 50.}
  \label{tab:user_study}
  \fontsize{9}{10.5}\selectfont
  \setlength{\tabcolsep}{3pt}
  \begin{tabular*}{\linewidth}{@{\extracolsep{\fill}}lrrrr@{}}
\toprule
Method & Task faithfulness & Physical plausibility & Visual quality & Overall preference \\
\midrule
% Generated from the read-only 2026-09-16 user-study refresh; Total only, original question order.
Text2Sim & \cellcolor[RGB]{198,221,211}$\mathbf{61.11}\,{\pm}\,2.59$ & \cellcolor[RGB]{193,218,207}$\mathbf{62.10}\,{\pm}\,2.59$ & \cellcolor[RGB]{204,225,215}$\mathbf{59.92}\,{\pm}\,3.01$ & \cellcolor[RGB]{197,221,209}$\mathbf{61.41}\,{\pm}\,3.11$ \\
GPT-5.6-Sol & \cellcolor[RGB]{255,255,255}50 (anchor) & \cellcolor[RGB]{255,255,255}50 (anchor) & \cellcolor[RGB]{255,255,255}50 (anchor) & \cellcolor[RGB]{255,255,255}50 (anchor) \\
Qwen-3.8-Max & \cellcolor[RGB]{252,246,247}$48.11\,{\pm}\,3.07$ & \cellcolor[RGB]{252,245,246}$47.82\,{\pm}\,3.41$ & \cellcolor[RGB]{248,233,235}$45.34\,{\pm}\,3.30$ & \cellcolor[RGB]{252,246,247}$48.02\,{\pm}\,3.57$ \\
Claude-Opus-5 & \cellcolor[RGB]{235,243,239}$53.97\,{\pm}\,2.97$ & \cellcolor[RGB]{240,246,243}$52.98\,{\pm}\,3.11$ & \cellcolor[RGB]{245,249,247}$51.98\,{\pm}\,3.19$ & \cellcolor[RGB]{234,242,238}$54.17\,{\pm}\,3.41$ \\
Code2Worlds & \cellcolor[RGB]{254,251,251}$49.17\,{\pm}\,2.76$ & \cellcolor[RGB]{247,231,233}$44.74\,{\pm}\,3.47$ & \cellcolor[RGB]{232,241,237}$54.56\,{\pm}\,3.10$ & \cellcolor[RGB]{252,247,248}$48.31\,{\pm}\,3.37$ \\
\bottomrule
\end{tabular*}

\end{table*}

% Place the qualitative Debug Cards comparison after the three main score tables.
\begin{figure*}[!t]
  \centering
  \includegraphics[width=\linewidth]{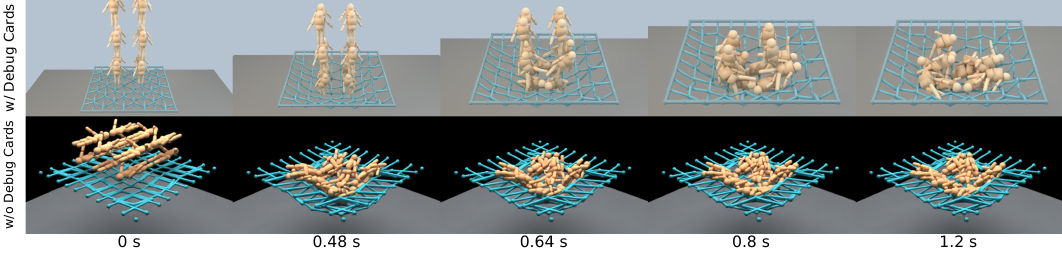}
  \caption{\textbf{Jointed-net Debug Cards comparison.} Full prompt: \emph{``Create a jointed net catching rigid body stacks demo: one 10*10-sized articulated rigid lattice made up of small capsule links and small sphere joints is hung above the ground with the four edges fixed. Gravity is enabled. Drop a compact stack of simple humanoids onto the center of the net so the lattice sags deeply, wrinkles, and settles.''} Rows show Text2Sim with and without Debug Cards, from top to bottom. Columns show shared video timestamps of 0, 0.48, 0.64, 0.8, and 1.2 seconds, from left to right.}
  \Description{Two rows of five early snapshots compare Text2Sim with and without Debug Cards. In the top row, a separated upright stack descends, its lower layer strikes the regular articulated net, and successive bodies rotate and pile as the net sags. The bottom row shows the flatter body arrangement contacting a net with folded-back strands. The first column shows the original initial rendered frames. Equal-height panels preserve image proportions, original camera views, and shared video timestamps.}
  \label{fig:debug_card_net}
\end{figure*}

\subsection{Debug Cards}
\label{sec:debug_cards}

\paragraph{Motivation.}
The agentic structure described above can decompose the complex task into specific subtasks and reduce the context burden on any single agent, but the agents still lack the physical expertise needed for authoring simulations. Therefore, we add Debug Cards to bridge this gap with simulation experience delivered through in-context learning~\citep{brown2020language}, connecting physical symptoms to diagnostic checks and repair strategies. This experience includes identifying initial interpenetration in an IPC case, checking energy injected by a controller, and comparing a reported success metric with the visible contact event. The key is to retain the diagnostic reasoning behind a repair in a form that can guide new scenes, rather than retain only the program that solved one demonstration.

Nevertheless, the amount of physical expertise which is potentially useful for any simulation task is extremely large. Thus, directly inserting all available experience into every prompt would consume the context that role decomposition is intended to preserve. To address this conflict, we instead organize this experience as a library of compact, human-readable \emph{Debug Cards}. Each card describes its conditions of use and associates symptoms with likely causes, distinguishing checks, and repair directions; it may also prohibit an apparently successful but physically invalid shortcut. Applicability tags identify physical modes, task attributes, failure signatures, and agent roles. The coupled choices in the comparison with and without Debug Cards (Fig.~\ref{fig:debug_card_net}), including a better net articulation structure, more reasonable initial ragdoll positions, and more suitable physical parameters for net bending, motivate guidance that connects an observed behavior to the component responsible for it.

\paragraph{Library construction.}
We construct the card library through iterative experience distillation from published graphics demonstrations. We collect demonstrations from SIGGRAPH and SIGGRAPH Asia technical papers together with their textual descriptions, emphasizing identifiable objects and actuation, observable contact or deformation, and outcomes that can be checked in execution records and rendered motion. These criteria make the demonstrations useful sources of diagnostic experience. An independent prompt-generation agent examines each demonstration and its source description, then writes a concise prompt describing the intended objects, interaction, actuation, and outcome. We run Text2Sim on these prompts to expose failures under the same pipeline used at inference time.

For each generated case, a separate evaluation agent compares intermediate programs, execution traces, and rendered results against the prompt and source demonstration. It identifies the failure and supporting evidence, then summarizes a validated diagnosis and repair procedure as Debug Cards. The affected cases are generated again with the new guidance. We repeat this process as new failure modes appear, editing or consolidating cards that express the same underlying issue. This distillation retains reusable checks and repair principles across demonstrations rather than prescribing a particular scene configuration.

\paragraph{Role-specific use.}
When utilizing the final Debug Cards library, our agentic structure now exhibits its compatibility with this library design. To be more specific, we let the Planner be the only agent with direct access to the card library. It considers cards compatible with the planned physical mode and agent role, then selects relevant guidance using the task and current execution evidence. Before each downstream call, it attaches the selected cards to that request. For example, in the choices illustrated in Fig.~\ref{fig:debug_card_net}, source-aware repair guidance belongs with the Planner's asset decisions, while initialization checks guide the Body Writer. A controller-diagnosis card can similarly guide the Action Writer and Critic. Downstream agents do not search the library independently. This separation lets the library accumulate broad experience while keeping downstream guidance focused and its use auditable.

% \section{Implementation Details}
% \label{sec:implementation}

\section{Evaluation and Results}
\label{sec:evaluation}

\subsection{Experimental Setup}
\label{sec:experiment_setup}

\begin{table}[t]
  \centering
  \caption{Current environment and implementation settings. GPU count describes the host.}
  \label{tab:implementation}
  \fontsize{8.5}{10}\selectfont
  \setlength{\tabcolsep}{3pt}
  \begin{tabularx}{\linewidth}{@{}lX@{}}
    \toprule
    Setting & Value \\
    \midrule
    Simulator & Genesis 0.4.5 \\
    Renderer & LuisaRender, commit \texttt{d176fd90} \\
    Runtime & Python 3.12.13; PyTorch 2.11.0; CUDA 12.8 \\
    GPUs & $8\times$ NVIDIA GeForce RTX 4090 D, 24\,GB each \\
    CPUs & $2\times$ AMD EPYC 7H12 (128 cores total)\\
    RAM & 512\,GiB \\
    Agent model & GPT-5.6-Sol \\
    Mesh generation & Meshy-5 \\
    Simulation budget & 30 attempts per case \\
    Simulation timeout & 3,600\,s per attempt \\
    \bottomrule
  \end{tabularx}
\end{table}

Tab.~\ref{tab:implementation} summarizes the current software environment, host hardware, and authoring settings. The Planner, Writers, and Critic share the listed language model; physical and rendering parameters are selected by agents for each case.

\paragraph{Tasks.}
We evaluate Text2Sim on 42 held-out text prompts drawn from published SIGGRAPH and SIGGRAPH Asia technical papers. The benchmark contains 10 rigid, 18 deformable, and 14 cloth cases. Each short prompt specifies the intended objects, physical interaction, actuation, and observable outcome without implementation details. To assess generalization beyond the source demonstrations used to construct Debug Cards and prevent information leakage, we split cases by source paper: once any case from a paper contributes to card extraction or refinement, no other case from that paper enters the test set. This protocol tests whether the extracted experience transfers to unseen cases from previously unseen source papers.

\paragraph{Baselines.}
We compare our method against four state-of-the-art baselines: GPT-5.6-Sol, Qwen-3.8-Max, and Claude-Opus-5 as general-purpose coding agents, and Code2Worlds~\citep{zhang2026code2worlds} as a simulation-authoring baseline. We provide the same task prompt and Genesis execution interface to these baselines, permit coding agents to use subagents, and cap simulation attempts at 30 per case, matching the settings of our method. We also provide these agents with Meshy API access, enabling them to generate the necessary digital assets for simulation, as our pipeline does.

% Register the comparison figures early to interleave them with the results discussion.
\begin{figure*}[!tp]
  \centering
  \includegraphics[width=\linewidth]{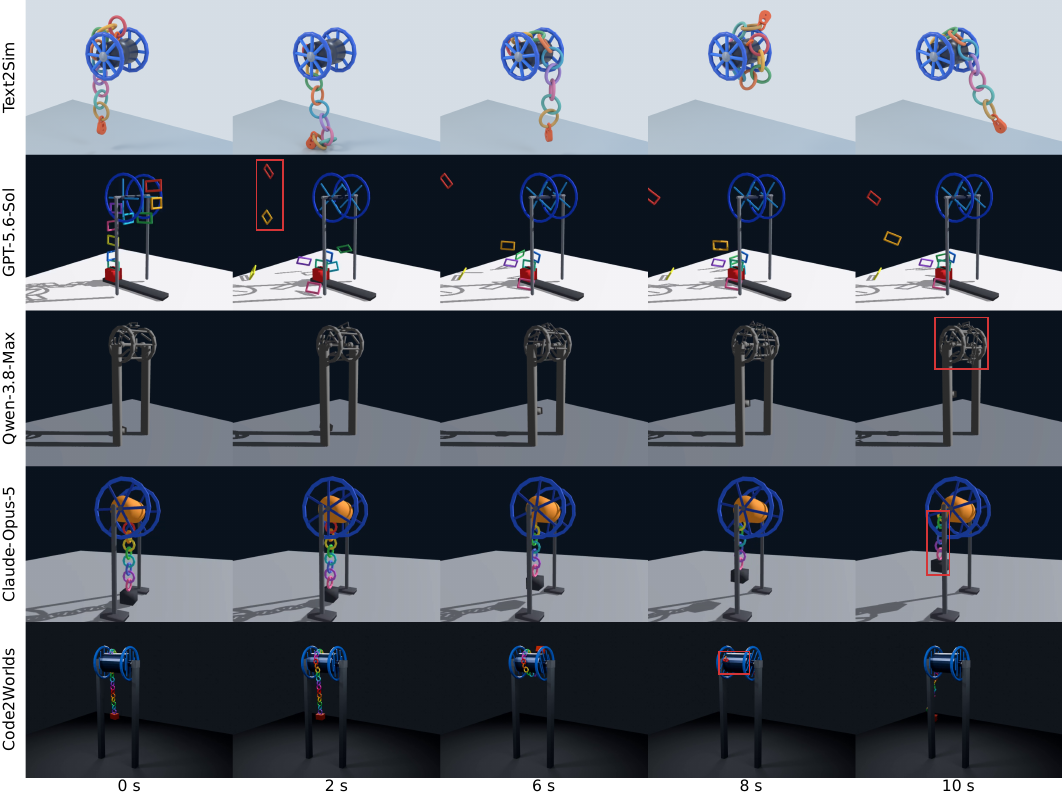}
  \caption{\textbf{Wheel-crawler task.} Full prompt: \emph{``Create a Cosserat-rod scene: a roller contains two large rigid wheel-like circular frames on either side and a central axle connecting them, allowing it to rotate along a fixed horizontal axis. A long chain containing several separate interlocked colorful rigid rings is connected to the roller, with the top link attached to the roller and the bottom link attached to a rigid anchor. Gravity is enabled. Drive the roller actuators consistently so the chain starts winding around the roller, lifting the anchor up. After fully rolled, the roller continues to roll, throwing the chain and the anchor out until the chain gets stretched again.''} Rows show Text2Sim, GPT-5.6-Sol, Qwen-3.8-Max, Claude-Opus-5, and Code2Worlds, from top to bottom. Columns show shared video timestamps of 0, 2, 6, 8, and 10 seconds, from left to right. Red boxes highlight regions discussed in Sec.~\ref{sec:main_results}. See the accompanying video for the complete motion sequences.}
  \Description{Five methods in a gapless grid, with columns synchronized at 0, 2, 6, 8, and 10 seconds. Text2Sim shows floor contact, lift, compact winding, and subsequent extension. GPT's links disperse. Qwen and Claude-Opus-5 show progressive anchor lifting. Code2Worlds shows winding near the drum followed by a chain extended downward. Red boxes mark dispersed GPT rings, poorly distinguished Qwen links, Claude's hanging anchor, and the Code2Worlds ring--drum contact region at 8 seconds.}
  \label{fig:cosserat_comparison}
\end{figure*}

\begin{figure*}[!tp]
  \centering
  \includegraphics[width=\linewidth]{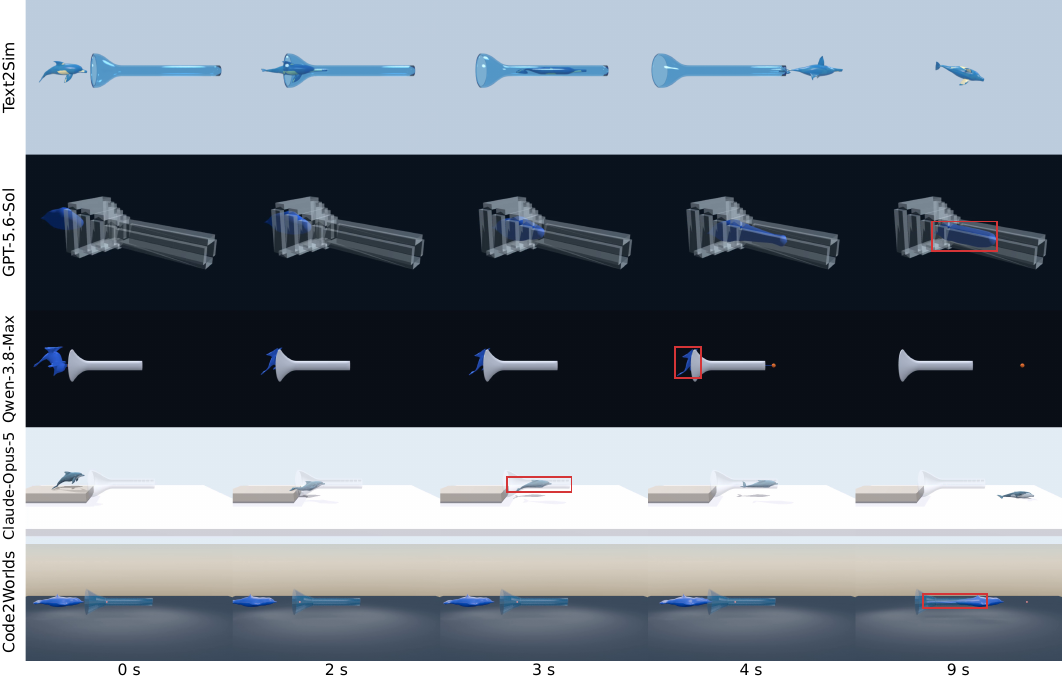}
  \caption{\textbf{Deformable dolphin task.} Full prompt: \emph{``Create a scene inspired by the dolphin-through-tube demo: a blue soft dolphin starts in front of a rigid funnel that narrows into a long thin horizontal tube. The diameter of the tube should be much smaller than the size of the dolphin. A strong force drags the front part of the dolphin through the funnel and tube, so the body elongates and squeezes through the narrow tube.''} Rows show Text2Sim, GPT-5.6-Sol, Qwen-3.8-Max, Claude-Opus-5, and Code2Worlds, from top to bottom. Columns show shared video timestamps of 0, 2, 3, 4, and 9 seconds, from left to right. Red boxes highlight regions discussed in Sec.~\ref{sec:main_results}. See the accompanying video for the complete motion sequences.}
  \Description{Five methods at five shared video times. Text2Sim is shown from the side with its original blue dolphin, transparent blue funnel and tube, and blue-gray background. The sequence shows the dolphin's elongation and passage, followed by a final view tracking its recovered compact shape after exit. The other rows retain their full original frames, showing different body shapes, funnel views, and degrees of passage at the same times. Red boxes mark GPT's body remaining in the tube, Qwen's body at the inlet, Claude's pale tube boundary, and Code2Worlds' elongated tail-to-body transition at 9 seconds.}
  \label{fig:dolphin_comparison}
\end{figure*}

\begin{figure*}[!tp]
  \centering
  \includegraphics[width=\linewidth]{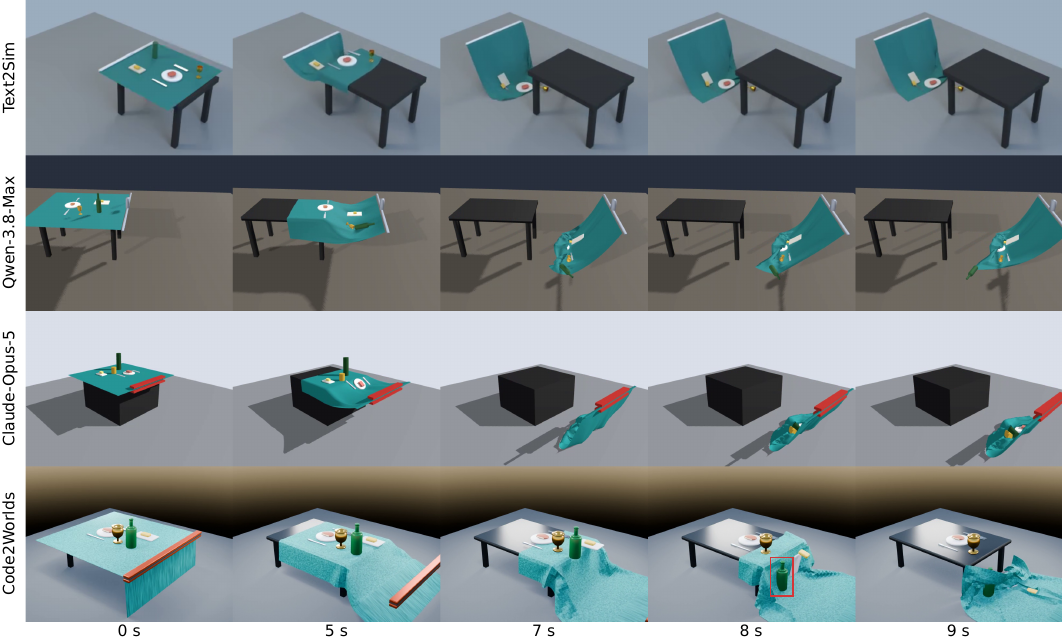}
  \caption{\textbf{Cloth slow-pull task.} Full prompt: \emph{``Create a scene inspired by a table-cloth slow-pull trick demo: one square tablecloth starts spread over a fixed rigid black table with one of the exposed edges attached to a moving clamp. Several simple rigid objects, including a white circular plate with a piece of pork on it, a knife and fork beside the plate separately, a plate with a piece of butter on it, a tall green wine bottle, and a gold wine cup, are placed on the tablecloth. Gravity is enabled. Move the clamp to pull the tablecloth so the cloth slides across the table and eventually falls away.''} Rows show Text2Sim, Qwen-3.8-Max, Claude-Opus-5, and Code2Worlds, from top to bottom. Columns show shared video timestamps of 0, 5, 7, 8, and 9 seconds, from left to right. GPT-5.6-Sol failed to execute this case and produced no completed video, so its row is omitted. Red boxes highlight regions discussed in Sec.~\ref{sec:main_results}. See the accompanying video for the complete motion sequences.}
  \Description{Four method rows for the tablecloth task: Text2Sim, Qwen-3.8-Max, Claude-Opus-5, and Code2Worlds. GPT-5.6-Sol failed on this case and is omitted. The available sequences show the cloth moving across the table, hanging over its edge, and reaching later configurations with different tabletop-object positions. A red box marks Code2Worlds' bottle--cloth contact region at 8 seconds.}
  \label{fig:tablecloth_comparison}
\end{figure*}

\subsection{Evaluation Metrics}
\label{sec:evaluation_metrics}

We use three complementary metrics: physical quality, visual quality, and human preference. Following decomposed video evaluation and fixed-rubric model judging~\citep{huang2024vbench,zheng2023llmjudge}, physical and visual scores use independent agent judgments conditioned on the prompt. Human preference scores come from a user study involving 60 participants from varied backgrounds. We evaluate all three metrics three times and report the standard deviation as an error estimate.

\paragraph{Physical quality.}
The physical score audits code, runtime records, and rendered evidence against prompt-specific requirements in four categories: Scene ($S$), Body ($B$), Action ($A$), and Render ($R$). Their 0--100 scores are combined before clamping and score caps as
\begin{equation}
S_{\mathrm{physical}} = w_SS + w_BB + w_AA + w_RR - P_{\mathrm{violation}}
\label{eq:physical_metric}
\end{equation}
Here we choose $w_S=0.20, w_B=w_A=0.275, w_R=0.25$ to emphasize the body and action components, which are more complex and more prone to errors. (However, in the later results, we find that this does not affect our method's lead over the baselines, since our method achieves the highest scores in all four categories.) \emph{Scene} checks the environment and layout, solver choice, gravity, boundaries, and global contact and friction settings. \emph{Body} checks object identities and counts, geometry and asset fidelity, colors and materials, rigid or deformable body types, and object-level physical parameters. \emph{Action} checks the requested actuation, collision and deformation behavior, event ordering, and final state, including whether motion follows physical simulation rather than teleportation or keyframes. \emph{Render} checks video availability and duration, camera coverage of the requested objects and events, lighting, material readability, and presentation quality. Its score combines rendering faithfulness (70\%) with aesthetic quality (30\%).

For Scene, Body, and Action, the evaluator assigns each atomic requirement an importance weight from 1 to 5 and a satisfaction value of 1, 0.5, or 0 for satisfied, partially satisfied or weakly evidenced, and absent or contradicted requirements, respectively; each category is their weighted mean scaled to 100. The penalty $P_{\mathrm{violation}}$ is 40 times the weighted fraction of explicit prohibitions that are violated, and is zero when no prohibitions apply. For example, replacing physical actuation with keyframes incurs this penalty when the prompt explicitly forbids that shortcut. After deduction, the score is clamped to $[0,100]$ and capped at 20 for unusable execution or rendering evidence, 40 for outputs consisting mainly of pre-rendered footage, keyframe animation, or visual effects instead of simulation, and 60 for a severe violation of an explicit prohibition. When several conditions apply, the lowest cap takes effect.

\paragraph{Visual quality.}
The visual score is isolated from implementation evidence: the evaluator sees only the prompt and the output video. It grades content fulfillment ($C$), demonstration clarity ($D$), visual polish ($P$), and SIGGRAPH demo appeal ($I$), with
\begin{equation}
S_{\mathrm{visual}} = w_CC + w_DD + w_PP + w_II.
\label{eq:visual_metric}
\end{equation}
Here we choose $w_C=w_D=0.3, w_P=0.25, w_I=0.15$ to emphasize content fulfillment and demonstration clarity, which are more important to the scene as a whole. (As noted in the physical-quality discussion, we observe that this choice does not affect our method's lead because it achieves the highest scores in all four dimensions.) \emph{Content fulfillment} ($C$) checks whether the requested objects, counts, appearance, spatial relationships, actions, and outcomes are visibly present. \emph{Demonstration clarity} ($D$) checks whether viewers can quickly identify the subject and target, follow the progression from the initial state through the interaction, and recognize the result. \emph{Visual polish} ($P$) checks geometry and material finish, lighting and shadows, separation of subjects from the background, camera composition, and visible defects such as clipping, broken geometry, noise, or distracting overlays. \emph{SIGGRAPH demo appeal} ($I$) checks deliberate visual focus, relevant richness without clutter, the impact of the demonstrated transformation, and watchability; it does not judge research novelty, code quality, or physical plausibility. These judgments use only visible evidence, without assuming that unobserved events occurred. If no valid generated video is available, all four visual dimensions and the Overall visual score are set to zero.

\paragraph{User-study metric.}
Blinded A/B comparisons assess the same 42 tasks.  We enrolled 60 participants, of whom 87\% have a STEM background, 74\% have LLM-driven agent experience, and 38\% have graphics knowledge. They view the task prompt and two outputs with method identities hidden, then answer four questions: \emph{task faithfulness}, ``Which result better matches the task description?''; \emph{physical plausibility}, ``Which result shows more physically plausible behavior?''; \emph{visual quality}, ``Which result is clearer, more visually complete, and more aesthetically pleasing?''; and \emph{overall preference}, ``Overall, which result do you prefer?'' Each question offers five ordered preference levels, from strongly preferring A through a tie to strongly preferring B, together with a ``Cannot judge'' option.

We select GPT-5.6-Sol as the anchor and fix its score at 50 as a common reference. Each remaining baseline---Qwen-3.8-Max, Claude-Opus-5, and Code2Worlds---is compared with both Text2Sim and the anchor; a direct Text2Sim--anchor comparison completes the seven pairings. Valid responses are oriented toward the evaluated method as $r\in\{-2,-1,0,1,2\}$, from strongly preferring its opponent to strongly preferring the method, and mapped to $50+25r$. Tab.~\ref{tab:user_study} uses each method's direct comparisons with the anchor, averaging valid responses within each case and then equally across cases, excluding ``Cannot judge'' responses. A tie maps to 50, and scores above or below 50 indicate preference relative to GPT-5.6-Sol. Each task--pair combination receives three prespecified ratings, distributed into assignments of 15 comparisons with no repeated task within an assignment. Trial order is randomized, and A/B positions alternate across repeats from a randomized initial ordering.

% Queue the ablation plot before Main Results so it lands beside the ablation analysis.
\begin{figure*}[!tp]
  \centering
  \includegraphics[width=\linewidth]{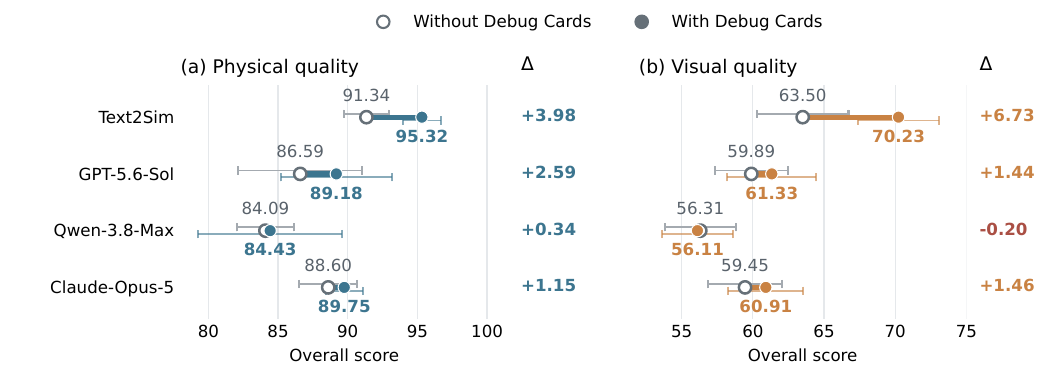}
  \caption{\textbf{Debug Card results and changes.} Paired points show Overall scores without cards (open) and with cards (filled), for (a) physical and (b) visual quality on 42 cases. Endpoint labels report the means; whiskers show within-case RMS scoring SD. Connecting segments and signed $\Delta$ labels emphasize With minus Without. The panels use different score ranges.}
  \Description{Paired absolute scores and signed changes for four pipelines. Text2Sim improves from 91.34 to 95.32 physically and from 63.50 to 70.23 visually. Qwen's visual score changes from 56.31 to 56.11; the other displayed changes are positive.}
  \label{fig:card_delta}
\end{figure*}

\subsection{Main Results}
\label{sec:main_results}

\paragraph{Physical quality.}
Tab.~\ref{tab:physical_results} reports the quantitative physical-quality results. Text2Sim obtains \OursPhysical{}, exceeding the strongest baseline, Claude-Opus-5 (88.60), by \PhysicalLead{} points. The three general-purpose agents obtain high Scene means (97.03--99.09), yet their larger gaps in Action and Render indicate difficulty in realizing the requested dynamic event beyond constructing its initial configuration. Relative to Claude-Opus-5, these gaps are 7.35 and 9.59 points, compared with 2.07 on Scene. Code2Worlds has the lowest Overall physical score, 67.91, trailing Text2Sim by 27.41 points. Its Scene and Action scores are 67.20 and 71.68, compared with Text2Sim's 99.10 and 96.10, showing weaknesses in both scene setup and event realization in the evaluated configuration.

The three example sequences illustrate the coupled requirements behind these scores; red boxes highlight regions discussed below. The accompanying video shows the complete motion sequences.

In the rigid wheel-crawler task (Fig.~\ref{fig:cosserat_comparison}), Text2Sim scores 95.74 physically and shows lower-link floor contact, anchor lift, compact winding, and renewed outward extension. GPT-5.6-Sol (67.27) has links disperse early while its anchor remains near the base. Qwen-3.8-Max (81.27) and Claude-Opus-5 (84.72) perform better physically, with correct links and anchor lifting, but they do not show the requested outward release, which is consistent with their physical scores being higher than GPT's but lower than ours. Code2Worlds (42.10) shows a critical physical failure with penetration of links into the roller, consistent with its low physical score.

In the deformable dolphin task (Fig.~\ref{fig:dolphin_comparison}), Text2Sim scores 98.13 and shows pronounced elongation inside the narrow tube and recovery of a compact shape after exit. GPT-5.6-Sol (71.03) produces visible squeezing but retains a stretched body inside the tube at 9 seconds. Qwen-3.8-Max (75.51) brings the nose through the tube while much of the body remains at the inlet; its execution records do not confirm tail exit. These partial successes correspond to their lower physical scores. Claude-Opus-5 (86.88) completes passage, although the pale tube makes the squeezing process less legible, which is consistent with its score being higher than those of the preceding baselines but lower than ours. Code2Worlds (39.85) presents a recognizable transit, but its code prescribes the principal translation and deformation using shape-key animation rather than physical forces (see the marked tail-to-body transition and the complete motion in the accompanying video), contributing to its poor physical score.

In the cloth task (Fig.~\ref{fig:tablecloth_comparison}), Text2Sim scores 96.16 and transports objects with the displaced cloth as it drapes over the table edge. Qwen-3.8-Max (91.02) and Claude-Opus-5 (88.91) also carry objects off the table, which is reflected in their scores being relatively close to ours. GPT-5.6-Sol (20.00) produces no completed video. Code2Worlds (69.08) leaves some tableware on the exposed tabletop; penetration also appears between the tablecloth and an upright bottle, making its score much lower than those of Qwen and Claude.

These coupled interactions motivate the complementary roles of our agentic structure and Debug Cards. The Planner coordinates Scene, Body, and Action Writers and routes execution evidence to an independent Critic, while retrieved cards supply simulation-specific checks and repair guidance for issues such as contact configuration, material choice, and actuation (Secs.~\ref{sec:multi_agent_structure} and~\ref{sec:debug_cards}). Text2Sim's higher Action scores and more complete event sequences support this combination of coordinated generation, critique, and reusable experience.

\paragraph{Visual quality.}
Tab.~\ref{tab:visual_results} reports the quantitative visual-quality results. Text2Sim obtains a visual score of \OursVisual{}, leading the strongest baseline, GPT-5.6-Sol (59.89), by \VisualLead{} points and achieving the largest mean in all four dimensions. Relative to GPT-5.6-Sol, the dimension gains are 13.22 points in demo appeal, 11.19 in visual polish, 9.96 in demonstration clarity, and 8.58 in content fulfillment.

Beyond realizing the physical events discussed above, the generated scenes must make those events easy and pleasing to inspect. In Fig.~\ref{fig:cosserat_comparison}, Text2Sim's visual score is 74.30, ahead of Claude-Opus-5 (61.58), Code2Worlds (59.70), Qwen-3.8-Max (44.18), and GPT-5.6-Sol (35.10). Text2Sim's contrasting link colors help viewers distinguish connected bodies, whereas Qwen-3.8-Max's largely monochrome, occluded links make their relationships harder to follow.

In Fig.~\ref{fig:dolphin_comparison}, Text2Sim scores 73.75 visually. The baseline scores are 69.70 for Code2Worlds, 57.15 for GPT-5.6-Sol, 49.28 for Claude-Opus-5, and 37.35 for Qwen-3.8-Max. Text2Sim's transparent tube exposes the deforming body throughout passage. Qwen-3.8-Max's opaque funnel conceals much of the interaction, while the pale tube in Claude-Opus-5 blends into the background. Code2Worlds makes the passage visually legible despite the prescribed motion discussed above, illustrating why the two metrics are complementary.

For the cloth task (Fig.~\ref{fig:tablecloth_comparison}), Text2Sim scores 75.40, compared with Code2Worlds (70.65), Qwen-3.8-Max (66.90), Claude-Opus-5 (66.75), and GPT-5.6-Sol (0.00, with no completed video). Text2Sim keeps the cloth, tabletop, and transported objects distinguishable through the pull. The richer surface appearance in Code2Worlds does not resolve the support ambiguity at the bottle--cloth contact highlighted at 8 seconds.

These presentation choices explain how visibility and visual organization affect demonstration clarity in addition to task completion. The larger gains in demo appeal and visual polish than in content fulfillment are consistent with this distinction. Assigning camera, lighting, and appearance to a dedicated Render Writer allows these aspects to be refined around an executed physical event, complementing the coordination and debugging process described above. The visual results support treating presentation as an explicit authoring objective within the overall pipeline.

% Keep the compact resource table near its discussion at the end of Main Results.
\begin{table}[!t]
  \centering
  \caption{Resource consumption statistics. Token counts are in millions; times are in minutes.}
  \label{tab:resources_consumption}
  \fontsize{9}{10.5}\selectfont
  \setlength{\tabcolsep}{3pt}
  \begin{tabular*}{\linewidth}{@{\extracolsep{\fill}}lrrr@{}}
\toprule
Method & \shortstack[r]{Tokens\\(M)} & \shortstack[r]{Time\\(min)} & \shortstack[r]{Time excl. GPU\\simulation (min)} \\
\midrule
Text2Sim & 29.1 & 101.4 & 58.0 \\
GPT-5.6-Sol & 24.8 & 60.9 & 41.9 \\
Qwen-3.8-Max & 23.9 & 95.1 & 55.8 \\
Claude-Opus-5 & 28.7 & 58.7 & 40.6 \\
Code2Worlds & 27.26 & 113.0 & 59.8 \\
\bottomrule
\end{tabular*}

\end{table}

% Allow the model comparison to reach the same page as its following subsection.
\begin{figure}[!t]
  \centering
  \includegraphics[width=\linewidth]{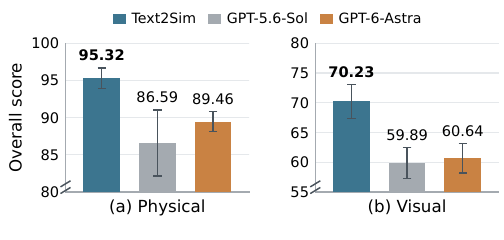}
  \caption{\textbf{Comparison with GPT-6-Astra.} Mean Overall scores with error bars over 42 cases for Text2Sim (blue), GPT-5.6-Sol (gray), and GPT-6-Astra (orange).}
  \Description{Two side-by-side panels each show three bars. The physical panel uses a truncated vertical axis from 80 to 100; the visual panel uses 55 to 80. Diagonal marks indicate truncation. Text2Sim scores 95.32 and 70.23, GPT-5.6-Sol scores 86.59 and 59.89, and GPT-6-Astra scores 89.46 and 60.64, respectively.}
  \label{fig:gpt6_comparison}
\end{figure}

% Queue the case comparison near the GPT-6-Astra discussion.
\begin{figure*}[!t]
  \centering
  \includegraphics[width=\linewidth]{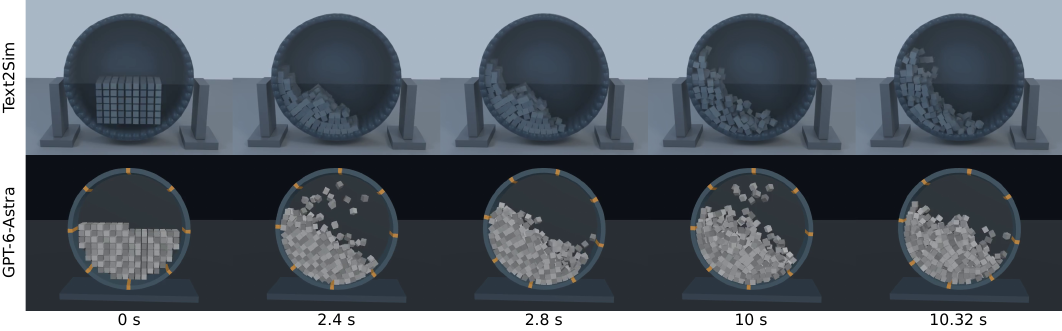}
  \caption{\textbf{Rigid tumbler comparison with GPT-6-Astra.} Full prompt: \emph{``Create a scene inspired by a rigid-body tumbler demo: hundreds of small gray rigid cubes start inside a hollow cylindrical drum with rigid circular side walls and a glass front wall. Gravity is enabled, and the drum is actuated with a slow continuous rotation. Friction lifts the bodies up, and then they avalanche down.''} Rows show Text2Sim and GPT-6-Astra, from top to bottom. Columns show shared video timestamps of 0, 2.4, 2.8, 10, and 10.32 seconds, from left to right. The paired early and late views show successive airborne and returning grains.}
  \Description{Two rows of five frames compare Text2Sim and GPT-6-Astra at shared video times. Text2Sim shows gray rigid blocks lifted along the drum wall and descending toward the bulk pile. GPT-6-Astra shows groups of blocks separated above the pile at 2.4 and 10 seconds and returning toward it at 2.8 and 10.32 seconds. The baseline retains its full original view; Text2Sim uses one fixed crop preserving the complete drum and its supports.}
  \label{fig:gpt6_tumbler_comparison}
\end{figure*}

% Queued early from results.tex to keep the application figures beside their discussion.
\begin{figure}[!t]
  \centering
  \includegraphics[width=\linewidth]{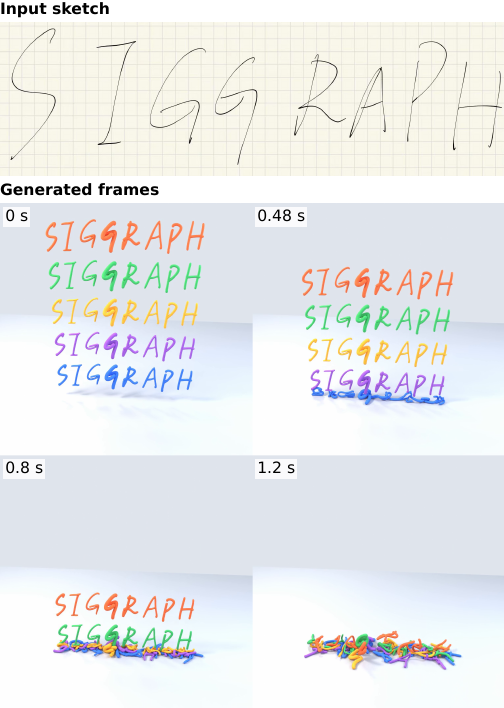}
  \caption{\textbf{Sketch-conditioned soft letters.} The handwritten ``SIGGRAPH'' input sketch is shown above four rendered frames of the generated deformable letters. Frames are arranged left to right, then top to bottom, at video times of 0, 0.48, 0.80, and 1.20 seconds. Full prompt: \emph{ ``Create a simulation of colorful soft letters (5 rows) falling, scattering, and settling into a loose pile.''}}
  \Description{A handwritten SIGGRAPH sketch above a two-by-two grid of full video frames at 0, 0.48, 0.8, and 1.2 seconds, read left to right and top to bottom. Initially, five colored rows of letters stand above a floor; later frames show lower rows bending against the floor, upper rows descending, and letters scattering.}
  \label{fig:sketch_to_simulation}
\end{figure}

% Queue the wide asset figure with the sketch so it reaches the multimodal discussion.
\begin{figure*}[!t]
  \centering
  \includegraphics[width=\linewidth]{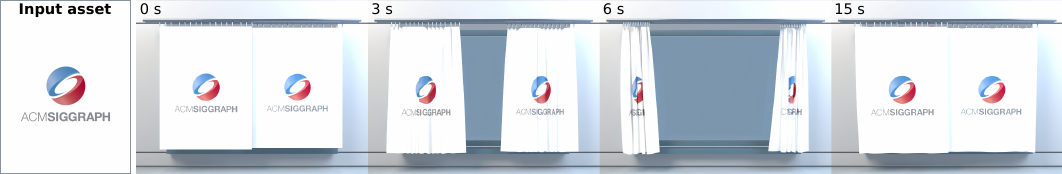}
  \caption{\textbf{Asset-conditioned curtains.} From left to right: the supplied cloth panel with its baked ACM SIGGRAPH texture, followed by four full frames of the generated scene at 0, 3, 6, and 15\,s. Full prompt: \emph{``Create a window curtain opening demo: two rectangular curtains bearing the supplied ACM SIGGRAPH graphic start hanging on two fixed horizontal rigid rods in front of a window, with their top edges attached to many small hooks along the rod. Gravity is enabled. Drive the slider hooks so the curtains first gather into dense folds at the sides and reveal the window, then slide back toward the center to stretch them again.''}}
  \Description{Five panels in one horizontal row. The leftmost panel shows a single white rectangular cloth asset with the ACM SIGGRAPH logo. The following four panels show two curtains initially closed at 0 seconds, opening at 3 seconds, gathered at the sides at 6 seconds, and closed again at 15 seconds. Both curtains carry the supplied logo, and all output panels retain the full camera view.}
  \label{fig:curtain_asset_to_simulation}
\end{figure*}

\paragraph{User preference.}
\label{sec:user_study}

Tab.~\ref{tab:user_study} reports the quantitative user-preference results, where scores of the anchor method (GPT-5.6-Sol) are set to 50 as a reference and higher scores indicate stronger user preference. Text2Sim leads all four aggregate user-study dimensions, obtaining 61.11 for task faithfulness, 62.10 for physical plausibility, 59.92 for visual quality, and 61.41 for overall preference. Its overall-preference score exceeds the strongest baseline, Claude-Opus-5 (54.17), by 7.24 points, with GPT-5.6-Sol fixed at 50. Code2Worlds receives 44.74 for physical plausibility and 54.56 for visual quality, while its overall preference is 48.31.

The human judgments connect the preceding analyses of event realization and presentation to perceived quality. Text2Sim's lead in both physical plausibility and visual quality indicates that its aggregate automatic-score advantage is also reflected in the user study. The contrasting physical and visual ratings for Code2Worlds further show why these dimensions should be assessed separately: favorable visual ratings can coexist with an overall preference below the anchor. This distinction is consistent with the differences in event realization discussed above, which appearance alone does not resolve. Across all five methods, the human physical-plausibility ranking matches the automatic physical-score ranking, while Text2Sim also ranks first under both visual evaluations. This agreement supports the relevance of the automatic criteria to human judgments at the aggregate level.

\paragraph{Resource consumption.}
\label{sec:resources_consumption}

Tab.~\ref{tab:resources_consumption} reports the resource consumption (token usage and generation time) of Text2Sim and four baselines. For both metrics, Text2Sim's resource consumption is of the same order of magnitude as that of the baselines. These results show that Text2Sim achieves high-quality results while requiring nearly the same level of resource consumption as these baselines, indicating the high efficiency of our pipeline design.

\subsection{Ablation Study}
\label{sec:ablation}

Text2Sim combines a hierarchy of agents with reusable simulation experience. The workflow coordinates construction, execution, and critique, while Debug Cards supply guidance for physical setup and repair. We conduct an ablation study to examine how these two elements contribute together. First, we compare the full Text2Sim pipeline with its variant without Debug Cards, assessing the value of retrieved experience within the agentic workflow. Second, we add Debug Cards to GPT-5.6-Sol, Qwen-3.8-Max, and Claude-Opus-5 to examine how the same form of guidance benefits other coding agents. Comparing Text2Sim with these agents under both card conditions further characterizes the performance of the coordinated pipeline. All variants are evaluated on the same 42 held-out tasks using the physical and visual metrics defined above.

Fig.~\ref{fig:card_delta} presents the paired overall scores and their changes for the four pipelines. Text2Sim improves from 91.34 to \OursPhysical{} in physical quality and from 63.50 to \OursVisual{} in visual quality when cards are included, gains of \CardPhysicalGain{} and \CardVisualGain{} points. These are the largest gains in both panels. GPT-5.6-Sol improves by 2.59 physical and 1.44 visual points, while Claude-Opus-5 improves by 1.15 and 1.46 points. Qwen-3.8-Max's physical score increases by 0.34 points. Text2Sim also achieves the highest absolute scores under both conditions. Without cards, it exceeds the strongest corresponding baseline by 2.74 physical and 3.61 visual points. With cards, its margins over the strongest card-augmented baselines are 5.57 and 8.90 points, respectively.

Fig.~\ref{fig:debug_card_net} visualizes one of these examples with the jointed-net task, showing a more regular net, a clearer falling stack, and deeper hammock deformation with Debug Cards. First, the source-aware repair card directs intrinsic mechanism defects back to asset generation. Together with progress diagnosis, this guides the replacement of a computationally expensive net topology with a tractable woven articulation. Second, initialization checks guide positive clearance between ragdolls; the agent's upright arrangement makes the individual bodies and their fall easy to distinguish. Third, asset-fidelity checks prompt a further revision of strand slack and passive hinge resistance, producing a deeper hammock under the load. Within the main run, the latter revision increases peak center sag from 0.427 to 0.661\,m. The cards supply transferable checks and repair directions, while the agents determine the case-specific geometry and parameters from execution evidence.

These results support the complementary roles of coordinated agents and reusable experience. Text2Sim's lead without cards shows that its construction and critique workflow already provides a strong basis for authoring physical events. The larger improvements when cards are introduced indicate that this workflow also makes effective use of the retrieved guidance. Within Text2Sim, cards improve Action and Render by 2.23 and 5.33 points, compared with 0.77 and 1.30 points for Scene and Body. All four visual dimensions improve, with the largest gains in demo appeal (9.21) and demonstration clarity (7.72). This pattern links the benefit of experience to carrying a scene through its intended interaction and presenting that interaction clearly. The positive physical gains across all four pipelines also demonstrate the usefulness of the distilled experience beyond Text2Sim. Together, the strongest absolute performance and the largest gains within Text2Sim support integrating Debug Cards with shared planning, specialized Writers, and execution-based critique, so that reusable guidance informs the component responsible for each construction or repair decision.

\subsection{Comparison with Newly Released GPT-6-Astra}
\label{sec:gpt6_comparison}

Following GPT-6-Astra's recent release, we conducted an additional comparison on the same 42 held-out prompts using our physical and visual metrics to evaluate GPT-6-Astra. Fig.~\ref{fig:gpt6_comparison} presents this result. As expected, GPT-6-Astra outperforms GPT-5.6-Sol on both metrics, demonstrating its improved capabilities. However, our method (whose base model is GPT-5.6-Sol) still leads GPT-6-Astra by 5.86 physical and 9.59 visual points. This result supports the value of our agentic pipeline design combined with skills extracted from high-quality simulations.

Fig.~\ref{fig:gpt6_tumbler_comparison} illustrates this comparison using the rigid tumbler case. Text2Sim (physical score 98.13 and visual score 72.83) shows bodies lifting along the inner wall and cascading toward the pile, while GPT-6-Astra (physical score 86.21 and visual score 62.45) exhibits repeated pronounced rebounds: groups of blocks bounce well above the bulk surface and then fall back, both early and late in the sequence. Text2Sim's more coherent bulk motion supports its higher physical score. Meanwhile, Text2Sim's more realistic front glass screen, with more realistic reflection and refraction, together with its clearer and more pleasing lighting, contributes to the higher visual score.

\section{Applications}
\label{sec:applications}

Text2Sim supports a broad range of downstream applications. Its executable, editable simulations expose geometry, material parameters, controls, and state trajectories alongside rendered videos, making these artifacts available for inspection, modification, and reuse. We select two representative applications to illustrate this broader scope: dataset construction and extension to multimodal input.

% Application figures are queued near the end of results.tex to avoid delayed floats.

\subsection{Dataset Construction}
\label{sec:dataset}

\paragraph{Data for vision--language models.}
Training vision--language models (VLMs) for dynamic scene understanding requires large amounts of video paired with informative descriptions. Web videos provide abundant observations, but their accompanying text often does not accurately describe the visible objects and interactions~\citep{wang2023internvid}. Preparing useful supervision therefore requires additional annotation, through manual labeling or automatic captioning with quality checks. For models that reason about 3D physical scenes, the challenge extends beyond description: videos do not directly expose geometry, material properties, actuation, or internal physical states, and recovering this dynamical representation from pixels alone is difficult.

Text2Sim addresses this data construction problem by enabling batch generation of aligned \emph{text description--video--dynamical representation} triples. Each item pairs the input request and rendered video with the generated program, asset identities, physical parameters, controls, and recorded states. The description and explicit dynamics are thus available alongside the visual observations. After checks of prompt fulfillment and simulation validity, these outputs can be curated into large collections of high-quality training and evaluation data that connect visible motion to its underlying physical structure.

We will release all 141 generated cases, including the 42 evaluation cases. Each case contains the prompt, rendered video, executable modules, assets, physical parameters, controls, and recorded states such as trajectories and contact events, along with Critic scores for curation. These attributes already cover what the applications below require, so building training sets, question sets, or benchmarks needs only selection and annotation, not new simulation authoring.

The recorded states also support more detailed supervision. State-derived annotations could describe object trajectories, event order, or whether a specified interaction occurs; questions could ask a VLM to identify the moving object, explain an observed outcome, or predict a subsequent state. Execution records would provide the basis for checking annotations in both training examples and evaluation questions.

The editable representation also supports controlled variants. Changing one material or actuation parameter and rerunning the program could produce paired examples for reasoning about physical interventions. Replaying a fixed trajectory with different rendering settings could test sensitivity to appearance while retaining the same motion. Such variants should stay in the same data split to prevent near-duplicate leakage. A curated collection would combine checks for asset integrity, finite and stable trajectories, prompt fulfillment, annotation consistency, and visual legibility with human review of ambiguous cases. The released programs can be modified and rerun directly to produce such variants.

\paragraph{Data for simulator benchmarks.}
Simulator benchmarking is another application of the generated data. Building a diverse collection of test scenes involves substantial expert effort: each setup requires geometry design, tuning of material parameters and actuation, and debugging to produce the intended physical interaction. Repeating this work across different contact configurations and material regimes makes benchmark construction labor-intensive. By generating, executing, and repairing scene programs from text requests in batches, Text2Sim provides a way to automate this preparation and expand the pool of editable candidate tests.

To form a benchmark, each generated case would fix geometry, materials, initial and boundary conditions, controls, duration, and the observables to be compared. Candidate tests include contact-heavy mechanisms, deformable settling, and cloth interactions. Runtime, memory, stability, and selected trajectory or contact quantities would be measured under documented discretizations and tolerances. Each comparison would specify compatible physical models and a validated analytic, experimental, or converged numerical reference, linking an editable scene to a reproducible benchmark configuration. The released cases provide such scenes in executable form, with logged trajectories and contact events as observables.

\subsection{Extension to Multimodal Input}
\label{sec:multimodal}

Text describes the desired physical event, while an additional input can specify object identity, shape, or appearance. We demonstrate two types of input: an image (a handwritten sketch) and a digital asset (a supplied textured cloth). Both enter the same construction, execution, critique, and repair workflow as text-only inputs, with simple supplementary instructions for their use. This extension provides users with an extra option to utilize various forms of input to specify scene details.

Fig.~\ref{fig:sketch_to_simulation} illustrates simulation from a handwritten sketch. The text specifies five rows of colorful soft letters and their falling motion, while the image supplies the glyph shapes and order without spelling them out in the prompt. The generated letters retain the sketch's distinctive forms and progress from orderly rows to bending, collisions, and scattering; the full video continues until they settle into a loose pile. This example shows how Text2Sim combines visual identity with the requested dynamics to turn a static drawing into an editable physical scene.

Fig.~\ref{fig:curtain_asset_to_simulation} demonstrates reuse of a supplied textured 3D asset. Both curtains retain the input's ACM SIGGRAPH graphic as they gather into folds at the sides, reveal the window, and return to a closed configuration. The printed graphics move and fold with the simulated cloth, while the agent supplies the surrounding window, support structure, and actuation. The result shows that Text2Sim can incorporate an existing asset's geometry and appearance into a new dynamic scene. Together, these examples demonstrate the flexibility of the shared authoring workflow in combining text, visual references, and reusable assets. The accompanying video shows the full motion sequences for both examples.

\section{Conclusions}
\label{sec:conclusion}
\label{sec:limitations}

We presented \textit{Text2Sim}, a pipeline that unifies asset preparation, physical setup, control, and rendering to turn text requests into executable, editable simulations. A Planner coordinates specialized Writers and an independent Critic, while retrieved Debug Cards bring experience from graphics demonstrations into construction and repair. We evaluate our method across various tasks, comparing it with four state-of-the-art baselines. These experiments show that Text2Sim achieves the highest aggregate physical, visual, and human-preference scores compared with four state-of-the-art baselines, demonstrating its robust and high-quality generation ability. Additional ablation studies support the design of the pipeline.

The broader potential of Text2Sim lies in making physical scenes easier to create, adapt, and reuse. Its explicit programs and recorded dynamics provide a basis for VLM training datasets and simulator benchmarks. The extension to multimodal input lets sketches and existing assets guide scene creation, connecting visual intent with executable behavior. Together, these applications point toward a common authoring interface for exploring physical scenarios and producing reusable content across graphics, learning, and simulation research.

The current system has two practical limitations. First, its physical scope is bounded by the material models, solvers, and contact mechanisms supported by the underlying simulator. Second, repeated agent calls, simulation, and rendering also impose time and computational costs, and when scenes scale up, these costs will increase rapidly. These two limitations leave room for future improvement and extensions of this work.

\FloatBarrier
\bibliographystyle{ACM-Reference-Format}
\bibliography{main}


\begin{thebibliography}{55}

%%% the Web address.

\ifx \showCODEN    \undefined \def \showCODEN     #1{\unskip}     \fi
\ifx \showISBNx    \undefined \def \showISBNx     #1{\unskip}     \fi
\ifx \showISBNxiii \undefined \def \showISBNxiii  #1{\unskip}     \fi
\ifx \showISSN     \undefined \def \showISSN      #1{\unskip}     \fi
\ifx \showLCCN     \undefined \def \showLCCN      #1{\unskip}     \fi
\ifx \shownote     \undefined \def \shownote      #1{#1}          \fi
\ifx \showarticletitle \undefined \def \showarticletitle #1{#1}   \fi
\ifx \showURL      \undefined \def \showURL       {\relax}        \fi
% invisible to TeX
\providecommand\bibfield[2]{#2}
\providecommand\bibinfo[2]{#2}
\providecommand\natexlab[1]{#1}
\providecommand\showeprint[2][]{arXiv:#2}

\bibitem[Bar-Tal et~al\mbox{.}(2024)]%
        {bar2024lumiere}
\bibfield{author}{\bibinfo{person}{Omer Bar-Tal}, \bibinfo{person}{Hila
  Chefer}, \bibinfo{person}{Omer Tov}, \bibinfo{person}{Charles Herrmann},
  \bibinfo{person}{Roni Paiss}, \bibinfo{person}{Shiran Zada},
  \bibinfo{person}{Ariel Ephrat}, \bibinfo{person}{Junhwa Hur},
  \bibinfo{person}{Guanghui Liu}, \bibinfo{person}{Amit Raj},
  \bibinfo{person}{Yuanzhen Li}, \bibinfo{person}{Michael Rubinstein},
  \bibinfo{person}{Tomer Michaeli}, \bibinfo{person}{Oliver Wang},
  \bibinfo{person}{Deqing Sun}, \bibinfo{person}{Tali Dekel}, {and}
  \bibinfo{person}{Inbar Mosseri}.} \bibinfo{year}{2024}\natexlab{}.
\newblock \showarticletitle{{Lumiere}: A Space-Time Diffusion Model for Video
  Generation}. In \bibinfo{booktitle}{\emph{SIGGRAPH Asia 2024 Conference
  Papers}}. \bibinfo{pages}{1--11}.
\newblock
\href{https://doi.org/10.1145/3680528.3687614}{doi:\nolinkurl{10.1145/3680528.3687614}}


\bibitem[Brown et~al\mbox{.}(2020)]%
        {brown2020language}
\bibfield{author}{\bibinfo{person}{Tom Brown}, \bibinfo{person}{Benjamin Mann},
  \bibinfo{person}{Nick Ryder}, \bibinfo{person}{Melanie Subbiah},
  \bibinfo{person}{Jared~D. Kaplan}, \bibinfo{person}{Prafulla Dhariwal},
  \bibinfo{person}{Arvind Neelakantan}, \bibinfo{person}{Pranav Shyam},
  \bibinfo{person}{Girish Sastry}, \bibinfo{person}{Amanda Askell},
  \bibinfo{person}{Sandhini Agarwal}, \bibinfo{person}{Ariel Herbert-Voss},
  \bibinfo{person}{Gretchen Krueger}, \bibinfo{person}{Tom Henighan},
  \bibinfo{person}{Rewon Child}, \bibinfo{person}{Aditya Ramesh},
  \bibinfo{person}{Daniel Ziegler}, \bibinfo{person}{Jeffrey Wu},
  \bibinfo{person}{Clemens Winter}, \bibinfo{person}{Chris Hesse},
  \bibinfo{person}{Mark Chen}, \bibinfo{person}{Eric Sigler},
  \bibinfo{person}{Mateusz Litwin}, \bibinfo{person}{Scott Gray},
  \bibinfo{person}{Benjamin Chess}, \bibinfo{person}{Jack Clark},
  \bibinfo{person}{Christopher Berner}, \bibinfo{person}{Sam McCandlish},
  \bibinfo{person}{Alec Radford}, \bibinfo{person}{Ilya Sutskever}, {and}
  \bibinfo{person}{Dario Amodei}.} \bibinfo{year}{2020}\natexlab{}.
\newblock \showarticletitle{Language models are few-shot learners}. In
  \bibinfo{booktitle}{\emph{Advances in Neural Information Processing
  Systems}}, Vol.~\bibinfo{volume}{33}. \bibinfo{pages}{1877--1901}.
\newblock


\bibitem[Chen et~al\mbox{.}(2026)]%
        {chen2026sam3d}
\bibfield{author}{\bibinfo{person}{Xingyu Chen}, \bibinfo{person}{Fu-Jen Chu},
  \bibinfo{person}{Pierre Gleize}, \bibinfo{person}{Kevin~J. Liang},
  \bibinfo{person}{Alexander Sax}, \bibinfo{person}{Hao Tang},
  \bibinfo{person}{Weiyao Wang}, \bibinfo{person}{Michelle Guo},
  \bibinfo{person}{Thibaut Hardin}, \bibinfo{person}{Xiang Li},
  \bibinfo{person}{Aohan Lin}, \bibinfo{person}{Jia-Wei Liu},
  \bibinfo{person}{Ziqi Ma}, \bibinfo{person}{Anushka Sagar},
  \bibinfo{person}{Bowen Song}, \bibinfo{person}{Xiaodong Wang},
  \bibinfo{person}{Jianing Yang}, \bibinfo{person}{Bowen Zhang},
  \bibinfo{person}{Piotr Doll{\'a}r}, \bibinfo{person}{Georgia Gkioxari},
  \bibinfo{person}{Matt Feiszli}, {and} \bibinfo{person}{Jitendra Malik}.}
  \bibinfo{year}{2026}\natexlab{}.
\newblock \showarticletitle{SAM 3D: 3Dfy Anything in Images}. In
  \bibinfo{booktitle}{\emph{Proceedings of the IEEE/CVF Conference on Computer
  Vision and Pattern Recognition}}.
\newblock


\bibitem[Chu et~al\mbox{.}(2021)]%
        {chu2021controls}
\bibfield{author}{\bibinfo{person}{Mengyu Chu}, \bibinfo{person}{Nils Thuerey},
  \bibinfo{person}{Hans-Peter Seidel}, \bibinfo{person}{Christian Theobalt},
  {and} \bibinfo{person}{Rhaleb Zayer}.} \bibinfo{year}{2021}\natexlab{}.
\newblock \showarticletitle{Learning Meaningful Controls for Fluids}.
\newblock \bibinfo{journal}{\emph{ACM Transactions on Graphics (TOG)}}
  \bibinfo{volume}{40}, \bibinfo{number}{4} (\bibinfo{year}{2021}),
  \bibinfo{pages}{100:1--100:13}.
\newblock
\href{https://doi.org/10.1145/3450626.3459845}{doi:\nolinkurl{10.1145/3450626.3459845}}


\bibitem[Coyne and Sproat(2001)]%
        {coyne2001wordseye}
\bibfield{author}{\bibinfo{person}{Bob Coyne} {and} \bibinfo{person}{Richard
  Sproat}.} \bibinfo{year}{2001}\natexlab{}.
\newblock \showarticletitle{{WordsEye}: An Automatic Text-to-Scene Conversion
  System}. In \bibinfo{booktitle}{\emph{Proceedings of the 28th annual
  conference on Computer graphics and interactive techniques}}.
  \bibinfo{pages}{487--496}.
\newblock
\href{https://doi.org/10.1145/383259.383316}{doi:\nolinkurl{10.1145/383259.383316}}


\bibitem[Ferguson et~al\mbox{.}(2021)]%
        {ferguson2021rigidipc}
\bibfield{author}{\bibinfo{person}{Zachary Ferguson}, \bibinfo{person}{Minchen
  Li}, \bibinfo{person}{Teseo Schneider}, \bibinfo{person}{Francisca
  Gil-Ureta}, \bibinfo{person}{Timothy Langlois}, \bibinfo{person}{Chenfanfu
  Jiang}, \bibinfo{person}{Denis Zorin}, \bibinfo{person}{Danny~M. Kaufman},
  {and} \bibinfo{person}{Daniele Panozzo}.} \bibinfo{year}{2021}\natexlab{}.
\newblock \showarticletitle{{Intersection-free Rigid Body Dynamics}}.
\newblock \bibinfo{journal}{\emph{ACM Transactions on Graphics (TOG)}}
  \bibinfo{volume}{40}, \bibinfo{number}{4}, Article \bibinfo{articleno}{183}
  (\bibinfo{year}{2021}).
\newblock
\href{https://doi.org/10.1145/3450626.3459802}{doi:\nolinkurl{10.1145/3450626.3459802}}


\bibitem[Ganeshan et~al\mbox{.}(2024)]%
        {ganeshan2024parsel}
\bibfield{author}{\bibinfo{person}{Aditya Ganeshan}, \bibinfo{person}{Ryan
  Huang}, \bibinfo{person}{Xianghao Xu}, \bibinfo{person}{R.~Kenny Jones},
  {and} \bibinfo{person}{Daniel Ritchie}.} \bibinfo{year}{2024}\natexlab{}.
\newblock \showarticletitle{{ParSEL}: Parameterized Shape Editing with
  Language}.
\newblock \bibinfo{journal}{\emph{ACM Transactions on Graphics (TOG)}}
  \bibinfo{volume}{43}, \bibinfo{number}{6} (\bibinfo{year}{2024}),
  \bibinfo{pages}{1--14}.
\newblock
\href{https://doi.org/10.1145/3687922}{doi:\nolinkurl{10.1145/3687922}}


\bibitem[Gao et~al\mbox{.}(2025)]%
        {gao2025seedance}
\bibfield{author}{\bibinfo{person}{Yu Gao}, \bibinfo{person}{Haoyuan Guo},
  \bibinfo{person}{Tuyen Hoang}, \bibinfo{person}{Weilin Huang},
  \bibinfo{person}{Lu Jiang}, \bibinfo{person}{Fangyuan Kong},
  \bibinfo{person}{Huixia Li}, \bibinfo{person}{Jiashi Li},
  \bibinfo{person}{Liang Li}, \bibinfo{person}{Xiaojie Li},
  \bibinfo{person}{Xunsong Li}, \bibinfo{person}{Yifu Li},
  \bibinfo{person}{Shanchuan Lin}, \bibinfo{person}{Zhijie Lin},
  \bibinfo{person}{Jiawei Liu}, \bibinfo{person}{Shu Liu},
  \bibinfo{person}{Xiaonan Nie}, \bibinfo{person}{Zhiwu Qing},
  \bibinfo{person}{Yuxi Ren}, \bibinfo{person}{Li Sun}, \bibinfo{person}{Zhi
  Tian}, \bibinfo{person}{Rui Wang}, \bibinfo{person}{Sen Wang},
  \bibinfo{person}{Guoqiang Wei}, \bibinfo{person}{Guohong Wu},
  \bibinfo{person}{Jie Wu}, \bibinfo{person}{Ruiqi Xia}, \bibinfo{person}{Fei
  Xiao}, \bibinfo{person}{Xuefeng Xiao}, \bibinfo{person}{Jiangqiao Yan},
  \bibinfo{person}{Ceyuan Yang}, \bibinfo{person}{Jianchao Yang},
  \bibinfo{person}{Runkai Yang}, \bibinfo{person}{Tao Yang},
  \bibinfo{person}{Yihang Yang}, \bibinfo{person}{Zilyu Ye},
  \bibinfo{person}{Xuejiao Zeng}, \bibinfo{person}{Yan Zeng},
  \bibinfo{person}{Heng Zhang}, \bibinfo{person}{Yang Zhao},
  \bibinfo{person}{Xiaozheng Zheng}, \bibinfo{person}{Peihao Zhu},
  \bibinfo{person}{Jiaxin Zou}, {and} \bibinfo{person}{Feilong Zuo}.}
  \bibinfo{year}{2025}\natexlab{}.
\newblock \showarticletitle{{Seedance 1.0}: Exploring the Boundaries of Video
  Generation Models}.
\newblock \bibinfo{journal}{\emph{arXiv preprint arXiv:2506.09113}}
  (\bibinfo{year}{2025}).
\newblock
\href{https://doi.org/10.48550/arXiv.2506.09113}{doi:\nolinkurl{10.48550/arXiv.2506.09113}}


\bibitem[{Genesis Authors}(2024)]%
        {Genesis}
\bibfield{author}{\bibinfo{person}{{Genesis Authors}}.}
  \bibinfo{year}{2024}\natexlab{}.
\newblock \bibinfo{title}{{Genesis}: A Generative and Universal Physics Engine
  for Robotics and Beyond}.
\newblock
\urldef\tempurl%
\url{https://github.com/Genesis-Embodied-AI/genesis-world}
\showURL{%
\tempurl}


\bibitem[Goel et~al\mbox{.}(2024)]%
        {goel2024motionediting}
\bibfield{author}{\bibinfo{person}{Purvi Goel}, \bibinfo{person}{Kuan-Chieh
  Wang}, \bibinfo{person}{C.~Karen Liu}, {and} \bibinfo{person}{Kayvon
  Fatahalian}.} \bibinfo{year}{2024}\natexlab{}.
\newblock \showarticletitle{Iterative Motion Editing with Natural Language}. In
  \bibinfo{booktitle}{\emph{ACM SIGGRAPH 2024 Conference Papers}}.
  \bibinfo{pages}{1--9}.
\newblock
\href{https://doi.org/10.1145/3641519.3657447}{doi:\nolinkurl{10.1145/3641519.3657447}}


\bibitem[Gumin et~al\mbox{.}(2025)]%
        {gumin2025procedural}
\bibfield{author}{\bibinfo{person}{Maxim Gumin}, \bibinfo{person}{Do~Heon Han},
  \bibinfo{person}{Seung~Jean Yoo}, \bibinfo{person}{Aditya Ganeshan},
  \bibinfo{person}{R.~Kenny Jones}, \bibinfo{person}{Kailiang Fu},
  \bibinfo{person}{Rio Aguina-Kang}, \bibinfo{person}{Stewart Morris}, {and}
  \bibinfo{person}{Daniel Ritchie}.} \bibinfo{year}{2025}\natexlab{}.
\newblock \showarticletitle{Procedural Scene Programs for Open-Universe Scene
  Generation: {LLM}-Free Error Correction via Program Search}. In
  \bibinfo{booktitle}{\emph{Proceedings of the SIGGRAPH Asia 2025 Conference
  Papers}}. \bibinfo{pages}{1--11}.
\newblock
\href{https://doi.org/10.1145/3757377.3763930}{doi:\nolinkurl{10.1145/3757377.3763930}}


\bibitem[Ho et~al\mbox{.}(2022)]%
        {ho2022videodiffusion}
\bibfield{author}{\bibinfo{person}{Jonathan Ho}, \bibinfo{person}{Tim
  Salimans}, \bibinfo{person}{Alexey Gritsenko}, \bibinfo{person}{William
  Chan}, \bibinfo{person}{Mohammad Norouzi}, {and} \bibinfo{person}{David~J.
  Fleet}.} \bibinfo{year}{2022}\natexlab{}.
\newblock \showarticletitle{{Video Diffusion Models}}. In
  \bibinfo{booktitle}{\emph{Neural Information Processing Systems (NeurIPS)}},
  Vol.~\bibinfo{volume}{35}. \bibinfo{pages}{8633--8646}.
\newblock
\urldef\tempurl%
\url{https://arxiv.org/abs/2204.03458}
\showURL{%
\tempurl}


\bibitem[H{\"o}llein et~al\mbox{.}(2023)]%
        {hollein2023text2room}
\bibfield{author}{\bibinfo{person}{Lukas H{\"o}llein}, \bibinfo{person}{Ang
  Cao}, \bibinfo{person}{Andrew Owens}, \bibinfo{person}{Justin Johnson}, {and}
  \bibinfo{person}{Matthias Nie{\ss}ner}.} \bibinfo{year}{2023}\natexlab{}.
\newblock \showarticletitle{Text2room: Extracting textured 3d meshes from 2d
  text-to-image models}. In \bibinfo{booktitle}{\emph{2023 IEEE/CVF
  International Conference on Computer Vision (ICCV)}}.
  \bibinfo{publisher}{IEEE}, \bibinfo{pages}{7875--7886}.
\newblock


\bibitem[Huang et~al\mbox{.}(2026)]%
        {huang2026learn2fold}
\bibfield{author}{\bibinfo{person}{Yanjia Huang}, \bibinfo{person}{Yunuo Chen},
  \bibinfo{person}{Ying Jiang}, \bibinfo{person}{Jinru Han},
  \bibinfo{person}{Zhengzhong Tu}, \bibinfo{person}{Yin Yang}, {and}
  \bibinfo{person}{Chenfanfu Jiang}.} \bibinfo{year}{2026}\natexlab{}.
\newblock \bibinfo{title}{Learn2Fold: Structured Origami Generation with World
  Model Planning}.
\newblock


\bibitem[Huang et~al\mbox{.}(2024)]%
        {huang2024vbench}
\bibfield{author}{\bibinfo{person}{Ziqi Huang}, \bibinfo{person}{Yinan He},
  \bibinfo{person}{Jiashuo Yu}, \bibinfo{person}{Fan Zhang},
  \bibinfo{person}{Chenyang Si}, \bibinfo{person}{Yuming Jiang},
  \bibinfo{person}{Yuanhan Zhang}, \bibinfo{person}{Tianxing Wu},
  \bibinfo{person}{Qingyang Jin}, \bibinfo{person}{Nattapol Chanpaisit},
  \bibinfo{person}{Yaohui Wang}, \bibinfo{person}{Xinyuan Chen},
  \bibinfo{person}{Limin Wang}, \bibinfo{person}{Dahua Lin},
  \bibinfo{person}{Yu Qiao}, {and} \bibinfo{person}{Ziwei Liu}.}
  \bibinfo{year}{2024}\natexlab{}.
\newblock \showarticletitle{VBench: Comprehensive Benchmark Suite for Video
  Generative Models}. In \bibinfo{booktitle}{\emph{IEEE Conference on Computer
  Vision and Pattern Recognition (CVPR)}}.
\newblock


\bibitem[Hui et~al\mbox{.}(2024)]%
        {hui2024qwen25coder}
\bibfield{author}{\bibinfo{person}{Binyuan Hui}, \bibinfo{person}{Jian Yang},
  \bibinfo{person}{Zeyu Cui}, \bibinfo{person}{Jiaxi Yang},
  \bibinfo{person}{Dayiheng Liu}, \bibinfo{person}{Lei Zhang},
  \bibinfo{person}{Tianyu Liu}, \bibinfo{person}{Jiajun Zhang},
  \bibinfo{person}{Bowen Yu}, \bibinfo{person}{Keming Lu}, \bibinfo{person}{Kai
  Dang}, \bibinfo{person}{Yang Fan}, \bibinfo{person}{Yichang Zhang},
  \bibinfo{person}{An Yang}, \bibinfo{person}{Rui Men}, \bibinfo{person}{Fei
  Huang}, \bibinfo{person}{Bo Zheng}, \bibinfo{person}{Yibo Miao},
  \bibinfo{person}{Shanghaoran Quan}, \bibinfo{person}{Yunlong Feng},
  \bibinfo{person}{Xingzhang Ren}, \bibinfo{person}{Xuancheng Ren},
  \bibinfo{person}{Jingren Zhou}, {and} \bibinfo{person}{Junyang Lin}.}
  \bibinfo{year}{2024}\natexlab{}.
\newblock \bibinfo{title}{Qwen2.5-Coder Technical Report}.
\newblock


\bibitem[Jiang et~al\mbox{.}(2016)]%
        {jiang2016material}
\bibfield{author}{\bibinfo{person}{Chenfanfu Jiang}, \bibinfo{person}{Craig
  Schroeder}, \bibinfo{person}{Joseph Teran}, \bibinfo{person}{Alexey
  Stomakhin}, {and} \bibinfo{person}{Andrew Selle}.}
  \bibinfo{year}{2016}\natexlab{}.
\newblock \showarticletitle{The material point method for simulating continuum
  materials}.
\newblock In \bibinfo{booktitle}{\emph{Acm siggraph 2016 courses}}.
  \bibinfo{pages}{1--52}.
\newblock


\bibitem[Juravsky et~al\mbox{.}(2022)]%
        {juravsky2022padl}
\bibfield{author}{\bibinfo{person}{Jordan Juravsky}, \bibinfo{person}{Yunrong
  Guo}, \bibinfo{person}{Sanja Fidler}, {and} \bibinfo{person}{Xue~Bin Peng}.}
  \bibinfo{year}{2022}\natexlab{}.
\newblock \showarticletitle{{PADL}: Language-Directed Physics-Based Character
  Control}. In \bibinfo{booktitle}{\emph{SIGGRAPH Asia 2022 Conference
  Papers}}. \bibinfo{pages}{1--9}.
\newblock
\href{https://doi.org/10.1145/3550469.3555391}{doi:\nolinkurl{10.1145/3550469.3555391}}


\bibitem[Konda and Tsitsiklis(1999)]%
        {konda1999actorcritic}
\bibfield{author}{\bibinfo{person}{Vijay Konda} {and} \bibinfo{person}{John
  Tsitsiklis}.} \bibinfo{year}{1999}\natexlab{}.
\newblock \showarticletitle{Actor-critic algorithms}.
\newblock \bibinfo{journal}{\emph{Advances in neural information processing
  systems}}  \bibinfo{volume}{12} (\bibinfo{year}{1999}).
\newblock


\bibitem[Koschier et~al\mbox{.}(2022)]%
        {koschier2022survey}
\bibfield{author}{\bibinfo{person}{Dan Koschier}, \bibinfo{person}{Jan Bender},
  \bibinfo{person}{Barbara Solenthaler}, {and} \bibinfo{person}{Matthias
  Teschner}.} \bibinfo{year}{2022}\natexlab{}.
\newblock \showarticletitle{A survey on SPH methods in computer graphics}. In
  \bibinfo{booktitle}{\emph{Computer graphics forum}},
  Vol.~\bibinfo{volume}{41}. Wiley Online Library, \bibinfo{pages}{737--760}.
\newblock


\bibitem[Li et~al\mbox{.}(2020)]%
        {li2020ipc}
\bibfield{author}{\bibinfo{person}{Minchen Li}, \bibinfo{person}{Zachary
  Ferguson}, \bibinfo{person}{Teseo Schneider}, \bibinfo{person}{Timothy
  Langlois}, \bibinfo{person}{Denis Zorin}, \bibinfo{person}{Daniele Panozzo},
  \bibinfo{person}{Chenfanfu Jiang}, {and} \bibinfo{person}{Danny~M. Kaufman}.}
  \bibinfo{year}{2020}\natexlab{}.
\newblock \showarticletitle{Incremental Potential Contact: Intersection- and
  Inversion-free, Large-Deformation Dynamics}.
\newblock \bibinfo{journal}{\emph{ACM Transactions on Graphics (TOG)}}
  \bibinfo{volume}{39}, \bibinfo{number}{4} (\bibinfo{year}{2020}),
  \bibinfo{pages}{49:1--49:20}.
\newblock
\href{https://doi.org/10.1145/3386569.3392425}{doi:\nolinkurl{10.1145/3386569.3392425}}


\bibitem[Li et~al\mbox{.}(2026)]%
        {li2026physics}
\bibfield{author}{\bibinfo{person}{Minchen Li}, \bibinfo{person}{Chenfanfu
  Jiang}, \bibinfo{person}{Zhaofeng Luo}, \bibinfo{person}{Wenxin Du},
  \bibinfo{person}{Chang Yu}, \bibinfo{person}{{\v{Z}}iga Kova{\v{c}}i{\v{c}}},
  {and} \bibinfo{person}{Tianyi Xie}.} \bibinfo{year}{2026}\natexlab{}.
\newblock \bibinfo{booktitle}{\emph{Physics-Based Simulation}}.
\newblock
\href{https://doi.org/10.5281/zenodo.20597655}{doi:\nolinkurl{10.5281/zenodo.20597655}}
\newblock
\shownote{Open-source online book. Live version available at
  \url{https://phys-sim-book.github.io/}}.


\bibitem[Li et~al\mbox{.}(2021)]%
        {li2021codimensional}
\bibfield{author}{\bibinfo{person}{Minchen Li}, \bibinfo{person}{Danny~M.
  Kaufman}, {and} \bibinfo{person}{Chenfanfu Jiang}.}
  \bibinfo{year}{2021}\natexlab{}.
\newblock \showarticletitle{Codimensional Incremental Potential Contact}.
\newblock \bibinfo{journal}{\emph{ACM Transactions on Graphics (TOG)}}
  \bibinfo{volume}{40}, \bibinfo{number}{4} (\bibinfo{year}{2021}),
  \bibinfo{pages}{170}.
\newblock


\bibitem[Liang et~al\mbox{.}(2023)]%
        {liang2023codepolicies}
\bibfield{author}{\bibinfo{person}{Jacky Liang}, \bibinfo{person}{Wenlong
  Huang}, \bibinfo{person}{Fei Xia}, \bibinfo{person}{Peng Xu},
  \bibinfo{person}{Karol Hausman}, \bibinfo{person}{Brian Ichter},
  \bibinfo{person}{Pete Florence}, {and} \bibinfo{person}{Andy Zeng}.}
  \bibinfo{year}{2023}\natexlab{}.
\newblock \showarticletitle{Code as Policies: Language Model Programs for
  Embodied Control}. In \bibinfo{booktitle}{\emph{2023 IEEE International
  Conference on Robotics and Automation (ICRA)}}. \bibinfo{publisher}{IEEE}.
\newblock


\bibitem[Lin et~al\mbox{.}(2026)]%
        {lin2026pat3d}
\bibfield{author}{\bibinfo{person}{Guying Lin}, \bibinfo{person}{Kemeng Huang},
  \bibinfo{person}{Michael Liu}, \bibinfo{person}{Ruihan Gao},
  \bibinfo{person}{Hanke Chen}, \bibinfo{person}{Lyuhao Chen},
  \bibinfo{person}{Beijia Lu}, \bibinfo{person}{Taku Komura},
  \bibinfo{person}{Yuan Liu}, \bibinfo{person}{Jun-Yan Zhu}, {and}
  \bibinfo{person}{Minchen Li}.} \bibinfo{year}{2026}\natexlab{}.
\newblock \showarticletitle{{PAT3D}: Physics-augmented text-to-3d scene
  generation}. In \bibinfo{booktitle}{\emph{International Conference on
  Learning Representations}}, Vol.~\bibinfo{volume}{2026}.
  \bibinfo{pages}{150281--150301}.
\newblock


\bibitem[Liu et~al\mbox{.}(2025)]%
        {liu2025ck}
\bibfield{author}{\bibinfo{person}{Michael Liu}, \bibinfo{person}{Xinlei Wang},
  {and} \bibinfo{person}{Minchen Li}.} \bibinfo{year}{2025}\natexlab{}.
\newblock \showarticletitle{Ck-mpm: A compact-kernel material point method}.
\newblock \bibinfo{journal}{\emph{ACM Transactions on Graphics (TOG)}}
  \bibinfo{volume}{44}, \bibinfo{number}{4} (\bibinfo{year}{2025}),
  \bibinfo{pages}{1--14}.
\newblock


\bibitem[Memery et~al\mbox{.}(2025)]%
        {memery2025cuetip}
\bibfield{author}{\bibinfo{person}{Sean Memery}, \bibinfo{person}{Kevin
  Denamganaï}, \bibinfo{person}{Jiaxin Zhang}, \bibinfo{person}{Zehai Tu},
  \bibinfo{person}{Yiwen Guo}, {and} \bibinfo{person}{Kartic Subr}.}
  \bibinfo{year}{2025}\natexlab{}.
\newblock \showarticletitle{{CueTip}: An Interactive and Explainable
  Physics-Aware Pool Assistant}. In \bibinfo{booktitle}{\emph{Proceedings of
  the Special Interest Group on Computer Graphics and Interactive Techniques
  Conference Conference Papers}}. \bibinfo{pages}{1--11}.
\newblock
\href{https://doi.org/10.1145/3721238.3730742}{doi:\nolinkurl{10.1145/3721238.3730742}}


\bibitem[{OpenAI}(2023)]%
        {openai2023gpt4}
\bibfield{author}{\bibinfo{person}{{OpenAI}}.} \bibinfo{year}{2023}\natexlab{}.
\newblock \bibinfo{title}{GPT-4 Technical Report}.
\newblock


\bibitem[Park et~al\mbox{.}(2026)]%
        {park2026mcpsim}
\bibfield{author}{\bibinfo{person}{Donggeun Park}, \bibinfo{person}{Hyeonbin
  Moon}, {and} \bibinfo{person}{Seunghwa Ryu}.}
  \bibinfo{year}{2026}\natexlab{}.
\newblock \showarticletitle{A Self-Correcting Multi-Agent LLM Framework for
  Language-Based Physics Simulation and Explanation}.
\newblock \bibinfo{journal}{\emph{npj Artificial Intelligence}}
  \bibinfo{volume}{2} (\bibinfo{year}{2026}), \bibinfo{pages}{10}.
\newblock


\bibitem[Pfaff et~al\mbox{.}(2026)]%
        {pfaff2026scenesmith}
\bibfield{author}{\bibinfo{person}{Nicholas Pfaff}, \bibinfo{person}{Thomas
  Cohn}, \bibinfo{person}{Sergey Zakharov}, \bibinfo{person}{Rick Cory}, {and}
  \bibinfo{person}{Russ Tedrake}.} \bibinfo{year}{2026}\natexlab{}.
\newblock \showarticletitle{Scenesmith: Agentic generation of simulation-ready
  indoor scenes}. In \bibinfo{booktitle}{\emph{Forty-third International
  Conference on Machine Learning}}.
\newblock


\bibitem[Poole et~al\mbox{.}(2022)]%
        {poole2022dreamfusion}
\bibfield{author}{\bibinfo{person}{Ben Poole}, \bibinfo{person}{Ajay Jain},
  \bibinfo{person}{Jonathan~T. Barron}, {and} \bibinfo{person}{Ben
  Mildenhall}.} \bibinfo{year}{2022}\natexlab{}.
\newblock \bibinfo{title}{DreamFusion: Text-to-3D using 2D Diffusion}.
\newblock


\bibitem[Pun et~al\mbox{.}(2025)]%
        {pun2025brickgpt}
\bibfield{author}{\bibinfo{person}{Ava Pun}, \bibinfo{person}{Kangle Deng},
  \bibinfo{person}{Ruixuan Liu}, \bibinfo{person}{Deva Ramanan},
  \bibinfo{person}{Changliu Liu}, {and} \bibinfo{person}{Jun-Yan Zhu}.}
  \bibinfo{year}{2025}\natexlab{}.
\newblock \showarticletitle{Generating physically stable and buildable brick
  structures from text}. In \bibinfo{booktitle}{\emph{2025 IEEE/CVF
  International Conference on Computer Vision (ICCV)}}.
  \bibinfo{publisher}{IEEE}, \bibinfo{pages}{14798--14809}.
\newblock


\bibitem[Schneider et~al\mbox{.}(2025)]%
        {schneider2025worldexplorer}
\bibfield{author}{\bibinfo{person}{Manuel-Andreas Schneider},
  \bibinfo{person}{Lukas Höllein}, {and} \bibinfo{person}{Matthias Nießner}.}
  \bibinfo{year}{2025}\natexlab{}.
\newblock \showarticletitle{{WorldExplorer}: Towards Generating Fully Navigable
  {3D} Scenes}. In \bibinfo{booktitle}{\emph{Proceedings of the SIGGRAPH Asia
  2025 Conference Papers}}. \bibinfo{pages}{1--11}.
\newblock
\href{https://doi.org/10.1145/3757377.3763946}{doi:\nolinkurl{10.1145/3757377.3763946}}


\bibitem[Shi et~al\mbox{.}(2025)]%
        {shi2025points2reward}
\bibfield{author}{\bibinfo{person}{Junyao Shi}, \bibinfo{person}{Joshua Smith},
  \bibinfo{person}{Jianing Qian}, {and} \bibinfo{person}{Dinesh Jayaraman}.}
  \bibinfo{year}{2025}\natexlab{}.
\newblock \showarticletitle{Points2Reward: Robotic Manipulation Rewards from
  Just One Video}. In \bibinfo{booktitle}{\emph{RSS 2025 Workshop on Semantic
  Reasoning and Goal Understanding in Robotics}}.
\newblock


\bibitem[Sifakis and Barbic(2012)]%
        {sifakis2012fem}
\bibfield{author}{\bibinfo{person}{Eftychios Sifakis} {and}
  \bibinfo{person}{Jernej Barbic}.} \bibinfo{year}{2012}\natexlab{}.
\newblock \showarticletitle{FEM simulation of 3D deformable solids: a
  practitioner's guide to theory, discretization and model reduction}.
\newblock In \bibinfo{booktitle}{\emph{Acm siggraph 2012 courses}}.
  \bibinfo{pages}{1--50}.
\newblock


\bibitem[Stomakhin et~al\mbox{.}(2013)]%
        {stomakhin2013snow}
\bibfield{author}{\bibinfo{person}{Alexey Stomakhin}, \bibinfo{person}{Craig
  Schroeder}, \bibinfo{person}{Lawrence Chai}, \bibinfo{person}{Joseph Teran},
  {and} \bibinfo{person}{Andrew Selle}.} \bibinfo{year}{2013}\natexlab{}.
\newblock \showarticletitle{A Material Point Method for Snow Simulation}.
\newblock \bibinfo{journal}{\emph{ACM Transactions on Graphics (TOG)}}
  \bibinfo{volume}{32}, \bibinfo{number}{4} (\bibinfo{year}{2013}),
  \bibinfo{pages}{1--10}.
\newblock
\href{https://doi.org/10.1145/2461912.2461948}{doi:\nolinkurl{10.1145/2461912.2461948}}


\bibitem[Sun et~al\mbox{.}(2024)]%
        {sun2024factorsim}
\bibfield{author}{\bibinfo{person}{Fan-Yun Sun}, \bibinfo{person}{S.~I.
  Harini}, \bibinfo{person}{Angela Yi}, \bibinfo{person}{Yihan Zhou},
  \bibinfo{person}{Alex Zook}, \bibinfo{person}{Jonathan Tremblay},
  \bibinfo{person}{Logan Cross}, \bibinfo{person}{Jiajun Wu}, {and}
  \bibinfo{person}{Nick Haber}.} \bibinfo{year}{2024}\natexlab{}.
\newblock \showarticletitle{FactorSim: Generative Simulation via Factorized
  Representation}. In \bibinfo{booktitle}{\emph{The Thirty-eighth Annual
  Conference on Neural Information Processing Systems}}.
\newblock


\bibitem[Tao et~al\mbox{.}(2024)]%
        {tao2024nirfs}
\bibfield{author}{\bibinfo{person}{Yuanyuan Tao}, \bibinfo{person}{Ivan
  Puhachov}, \bibinfo{person}{Derek Nowrouzezahrai}, {and}
  \bibinfo{person}{Paul Kry}.} \bibinfo{year}{2024}\natexlab{}.
\newblock \showarticletitle{Neural Implicit Reduced Fluid Simulation}. In
  \bibinfo{booktitle}{\emph{SIGGRAPH Asia 2024 Conference Papers}}.
  \bibinfo{pages}{1--11}.
\newblock


\bibitem[Wang et~al\mbox{.}(2026a)]%
        {wang2026chronollm}
\bibfield{author}{\bibinfo{person}{Jingquan Wang}, \bibinfo{person}{Andrew
  Negrut}, \bibinfo{person}{Harry Zhang}, \bibinfo{person}{Khailanii Slaton},
  \bibinfo{person}{Shu Wang}, \bibinfo{person}{Radu Serban},
  \bibinfo{person}{Jinlong Wu}, {and} \bibinfo{person}{Dan Negrut}.}
  \bibinfo{year}{2026}\natexlab{a}.
\newblock \showarticletitle{ChronoLLM: Customizing Language Models for
  Physics-Based Simulation Code Generation}.
\newblock \bibinfo{journal}{\emph{Multibody System Dynamics}}
  (\bibinfo{year}{2026}).
\newblock


\bibitem[Wang et~al\mbox{.}(2023b)]%
        {wang2024gensim}
\bibfield{author}{\bibinfo{person}{Lirui Wang}, \bibinfo{person}{Yiyang Ling},
  \bibinfo{person}{Zhecheng Yuan}, \bibinfo{person}{Mohit Shridhar},
  \bibinfo{person}{Chen Bao}, \bibinfo{person}{Yuzhe Qin},
  \bibinfo{person}{Bailin Wang}, \bibinfo{person}{Huazhe Xu}, {and}
  \bibinfo{person}{Xiaolong Wang}.} \bibinfo{year}{2023}\natexlab{b}.
\newblock \showarticletitle{GenSim: Generating Robotic Simulation Tasks via
  Large Language Models}.
\newblock \bibinfo{journal}{\emph{International Conference on Learning
  Representations (ICLR), 2024}} (\bibinfo{year}{2023}).
\newblock


\bibitem[Wang et~al\mbox{.}(2023a)]%
        {wang2023internvid}
\bibfield{author}{\bibinfo{person}{Yi Wang}, \bibinfo{person}{Yinan He},
  \bibinfo{person}{Yizhuo Li}, \bibinfo{person}{Kunchang Li},
  \bibinfo{person}{Jiashuo Yu}, \bibinfo{person}{Xin Ma},
  \bibinfo{person}{Xinhao Li}, \bibinfo{person}{Guo Chen},
  \bibinfo{person}{Xinyuan Chen}, \bibinfo{person}{Yaohui Wang},
  \bibinfo{person}{Conghui He}, \bibinfo{person}{Ping Luo},
  \bibinfo{person}{Ziwei Liu}, \bibinfo{person}{Yali Wang},
  \bibinfo{person}{Limin Wang}, {and} \bibinfo{person}{Yu Qiao}.}
  \bibinfo{year}{2023}\natexlab{a}.
\newblock \showarticletitle{{InternVid}: A Large-scale Video-Text Dataset for
  Multimodal Understanding and Generation}.
\newblock \bibinfo{journal}{\emph{arXiv preprint arXiv:2307.06942}}
  (\bibinfo{year}{2023}).
\newblock
\urldef\tempurl%
\url{https://arxiv.org/abs/2307.06942}
\showURL{%
\tempurl}


\bibitem[Wang et~al\mbox{.}(2026b)]%
        {wang2026simuscene}
\bibfield{author}{\bibinfo{person}{Yanan Wang}, \bibinfo{person}{Renxi Wang},
  \bibinfo{person}{Yongxin Wang}, \bibinfo{person}{Xuezhi Liang},
  \bibinfo{person}{Fajri Koto}, \bibinfo{person}{Timothy Baldwin},
  \bibinfo{person}{Xiaodan Liang}, {and} \bibinfo{person}{Haonan Li}.}
  \bibinfo{year}{2026}\natexlab{b}.
\newblock \showarticletitle{SimuScene: Training and Benchmarking Code
  Generation to Simulate Physical Scenarios}.
\newblock \bibinfo{journal}{\emph{arXiv preprint arXiv:2602.10840}}
  (\bibinfo{year}{2026}).
\newblock


\bibitem[Wang et~al\mbox{.}(2023c)]%
        {wang2024robogen}
\bibfield{author}{\bibinfo{person}{Yufei Wang}, \bibinfo{person}{Zhou Xian},
  \bibinfo{person}{Feng Chen}, \bibinfo{person}{Tsun-Hsuan Wang},
  \bibinfo{person}{Yian Wang}, \bibinfo{person}{Katerina Fragkiadaki},
  \bibinfo{person}{Zackory Erickson}, \bibinfo{person}{David Held}, {and}
  \bibinfo{person}{Chuang Gan}.} \bibinfo{year}{2023}\natexlab{c}.
\newblock \showarticletitle{RoboGen: Towards Unleashing Infinite Data for
  Automated Robot Learning via Generative Simulation}.
\newblock \bibinfo{journal}{\emph{ICML 2024}} (\bibinfo{year}{2023}).
\newblock


\bibitem[Wu et~al\mbox{.}(2023)]%
        {wu2023autogen}
\bibfield{author}{\bibinfo{person}{Qingyun Wu}, \bibinfo{person}{Gagan Bansal},
  \bibinfo{person}{Jieyu Zhang}, \bibinfo{person}{Yiran Wu},
  \bibinfo{person}{Beibin Li}, \bibinfo{person}{Erkang Zhu},
  \bibinfo{person}{Li Jiang}, \bibinfo{person}{Xiaoyun Zhang},
  \bibinfo{person}{Shaokun Zhang}, \bibinfo{person}{Jiale Liu},
  \bibinfo{person}{Ahmed~Hassan Awadallah}, \bibinfo{person}{Ryen~W White},
  \bibinfo{person}{Doug Burger}, {and} \bibinfo{person}{Chi Wang}.}
  \bibinfo{year}{2023}\natexlab{}.
\newblock \showarticletitle{{AutoGen}: Enabling Next-Gen {LLM} Applications via
  Multi-Agent Conversation}.
\newblock \bibinfo{journal}{\emph{arXiv preprint arXiv:2308.08155}}
  (\bibinfo{year}{2023}).
\newblock
\href{https://doi.org/10.48550/arXiv.2308.08155}{doi:\nolinkurl{10.48550/arXiv.2308.08155}}


\bibitem[Xia et~al\mbox{.}(2026)]%
        {xia2026sage}
\bibfield{author}{\bibinfo{person}{Hongchi Xia}, \bibinfo{person}{Xuan Li},
  \bibinfo{person}{Zhaoshuo Li}, \bibinfo{person}{Qianli Ma},
  \bibinfo{person}{Jiashu Xu}, \bibinfo{person}{Ming-Yu Liu},
  \bibinfo{person}{Yin Cui}, \bibinfo{person}{Tsung-Yi Lin},
  \bibinfo{person}{Wei-Chiu Ma}, \bibinfo{person}{Shenlong Wang},
  \bibinfo{person}{Shuran Song}, {and} \bibinfo{person}{Fangyin Wei}.}
  \bibinfo{year}{2026}\natexlab{}.
\newblock \showarticletitle{SAGE: Scalable Agentic 3D Scene Generation for
  Embodied AI}.
\newblock \bibinfo{journal}{\emph{arXiv preprint arXiv:2602.10116}}
  (\bibinfo{year}{2026}).
\newblock


\bibitem[Yang et~al\mbox{.}(2024a)]%
        {yang2024sweagent}
\bibfield{author}{\bibinfo{person}{John Yang}, \bibinfo{person}{Carlos
  Jimenez}, \bibinfo{person}{Alexander Wettig}, \bibinfo{person}{Kilian
  Lieret}, \bibinfo{person}{Shunyu Yao}, \bibinfo{person}{Karthik Narasimhan},
  {and} \bibinfo{person}{Ofir Press}.} \bibinfo{year}{2024}\natexlab{a}.
\newblock \showarticletitle{Swe-agent: Agent-computer interfaces enable
  automated software engineering}.
\newblock \bibinfo{journal}{\emph{Advances in Neural Information Processing
  Systems}}  \bibinfo{volume}{37} (\bibinfo{year}{2024}),
  \bibinfo{pages}{50528--50652}.
\newblock


\bibitem[Yang et~al\mbox{.}(2024b)]%
        {yang2024holodeck}
\bibfield{author}{\bibinfo{person}{Yue Yang}, \bibinfo{person}{Fan-Yun Sun},
  \bibinfo{person}{Luca Weihs}, \bibinfo{person}{Eli VanderBilt},
  \bibinfo{person}{Alvaro Herrasti}, \bibinfo{person}{Winson Han},
  \bibinfo{person}{Jiajun Wu}, \bibinfo{person}{Nick Haber},
  \bibinfo{person}{Ranjay Krishna}, \bibinfo{person}{Lingjie Liu},
  \bibinfo{person}{Chris Callison-Burch}, \bibinfo{person}{Mark Yatskar},
  \bibinfo{person}{Aniruddha Kembhavi}, {and} \bibinfo{person}{Christopher
  Clark}.} \bibinfo{year}{2024}\natexlab{b}.
\newblock \showarticletitle{Holodeck: Language guided generation of 3d embodied
  ai environments}. In \bibinfo{booktitle}{\emph{CVPR 2024}}.
\newblock


\bibitem[Yao et~al\mbox{.}(2022)]%
        {yao2022react}
\bibfield{author}{\bibinfo{person}{Shunyu Yao}, \bibinfo{person}{Jeffrey Zhao},
  \bibinfo{person}{Dian Yu}, \bibinfo{person}{Nan Du}, \bibinfo{person}{Izhak
  Shafran}, \bibinfo{person}{Karthik Narasimhan}, {and} \bibinfo{person}{Yuan
  Cao}.} \bibinfo{year}{2022}\natexlab{}.
\newblock \showarticletitle{ReAct: Synergizing reasoning and acting in language
  models}. In \bibinfo{booktitle}{\emph{ICLR 2023}}.
\newblock


\bibitem[Zhang et~al\mbox{.}(2026a)]%
        {zhang2026gsagent}
\bibfield{author}{\bibinfo{person}{Hongxin Zhang}, \bibinfo{person}{Chunru
  Lin}, \bibinfo{person}{Junyan Li}, \bibinfo{person}{Zhou Xian},
  \bibinfo{person}{Tsun-Hsuan Wang}, {and} \bibinfo{person}{Chuang Gan}.}
  \bibinfo{year}{2026}\natexlab{a}.
\newblock \showarticletitle{{GS-Agent}: Creating {4D} Physical Worlds With
  Generative Simulation}.
\newblock \bibinfo{journal}{\emph{arXiv preprint arXiv:2607.21522}}
  (\bibinfo{year}{2026}).
\newblock
\href{https://doi.org/10.48550/arXiv.2607.21522}{doi:\nolinkurl{10.48550/arXiv.2607.21522}}


\bibitem[Zhang et~al\mbox{.}(2026b)]%
        {zhang2026code2worlds}
\bibfield{author}{\bibinfo{person}{Yi Zhang}, \bibinfo{person}{Yunshuang Wang},
  \bibinfo{person}{Zeyu Zhang}, {and} \bibinfo{person}{Hao Tang}.}
  \bibinfo{year}{2026}\natexlab{b}.
\newblock \showarticletitle{Code2Worlds: Empowering Coding LLMs for 4D World
  Generation}. In \bibinfo{booktitle}{\emph{Forty-third International
  Conference on Machine Learning}}.
\newblock


\bibitem[Zhao et~al\mbox{.}(2016)]%
        {zhao2016relationship}
\bibfield{author}{\bibinfo{person}{Xi Zhao}, \bibinfo{person}{Ruizhen Hu},
  \bibinfo{person}{Paul Guerrero}, \bibinfo{person}{Niloy Mitra}, {and}
  \bibinfo{person}{Taku Komura}.} \bibinfo{year}{2016}\natexlab{}.
\newblock \showarticletitle{Relationship Templates for Creating Scene
  Variations}.
\newblock \bibinfo{journal}{\emph{ACM Transactions on Graphics (TOG)}}
  \bibinfo{volume}{35}, \bibinfo{number}{6} (\bibinfo{year}{2016}),
  \bibinfo{pages}{1--13}.
\newblock
\href{https://doi.org/10.1145/2980179.2982410}{doi:\nolinkurl{10.1145/2980179.2982410}}


\bibitem[Zheng et~al\mbox{.}(2026)]%
        {zheng2026barrierfree}
\bibfield{author}{\bibinfo{person}{Juntian Zheng}, \bibinfo{person}{Zhaofeng
  Luo}, {and} \bibinfo{person}{Minchen Li}.} \bibinfo{year}{2026}\natexlab{}.
\newblock \showarticletitle{Robust and Efficient Penetration-Free
  Elastodynamics without Barriers}.
\newblock \bibinfo{journal}{\emph{ACM Transactions on Graphics}}
  \bibinfo{volume}{45}, \bibinfo{number}{5} (\bibinfo{year}{2026}),
  \bibinfo{pages}{1--20}.
\newblock
\href{https://doi.org/10.1145/3811035}{doi:\nolinkurl{10.1145/3811035}}


\bibitem[Zheng et~al\mbox{.}(2023)]%
        {zheng2023llmjudge}
\bibfield{author}{\bibinfo{person}{Lianmin Zheng}, \bibinfo{person}{Wei-Lin
  Chiang}, \bibinfo{person}{Ying Sheng}, \bibinfo{person}{Siyuan Zhuang},
  \bibinfo{person}{Zhanghao Wu}, \bibinfo{person}{Yonghao Zhuang},
  \bibinfo{person}{Zi Lin}, \bibinfo{person}{Zhuohan Li},
  \bibinfo{person}{Dacheng Li}, \bibinfo{person}{Eric Xing},
  \bibinfo{person}{Hao Zhang}, \bibinfo{person}{Joseph Gonzalez}, {and}
  \bibinfo{person}{Ion Stoica}.} \bibinfo{year}{2023}\natexlab{}.
\newblock \showarticletitle{Judging LLM-as-a-Judge with MT-Bench and Chatbot
  Arena}. In \bibinfo{booktitle}{\emph{Advances in Neural Information
  Processing Systems}}.
\newblock


\bibitem[Zheng et~al\mbox{.}(2022)]%
        {zheng2022luisarender}
\bibfield{author}{\bibinfo{person}{Shaokun Zheng}, \bibinfo{person}{Zhiqian
  Zhou}, \bibinfo{person}{Xin Chen}, \bibinfo{person}{Difei Yan},
  \bibinfo{person}{Chuyan Zhang}, \bibinfo{person}{Yuefeng Geng},
  \bibinfo{person}{Yan Gu}, {and} \bibinfo{person}{Kun Xu}.}
  \bibinfo{year}{2022}\natexlab{}.
\newblock \showarticletitle{Luisarender: A high-performance rendering framework
  with layered and unified interfaces on stream architectures}.
\newblock \bibinfo{journal}{\emph{ACM Transactions on Graphics (TOG)}}
  \bibinfo{volume}{41}, \bibinfo{number}{6} (\bibinfo{year}{2022}),
  \bibinfo{pages}{1--19}.
\newblock
\href{https://doi.org/10.1145/3550454.3555463}{doi:\nolinkurl{10.1145/3550454.3555463}}


\bibitem[Zhu et~al\mbox{.}(2026)]%
        {zhu2026relaxflow}
\bibfield{author}{\bibinfo{person}{Jiayin Zhu}, \bibinfo{person}{Guoji Fu},
  \bibinfo{person}{Xiaolu Liu}, \bibinfo{person}{Qiyuan He},
  \bibinfo{person}{Yicong Li}, {and} \bibinfo{person}{Angela Yao}.}
  \bibinfo{year}{2026}\natexlab{}.
\newblock \showarticletitle{RelaxFlow: Text-Driven Amodal 3D Generation}.
\newblock \bibinfo{journal}{\emph{International Conference on Machine Learning
  (ICML'26 Spotlight)}} (\bibinfo{year}{2026}).
\newblock


\end{thebibliography}

% \appendix
% \input{text/appendix}

\end{document}